\documentclass[twocolumn]{aastex62}

\def\lapp{\ifmmode\stackrel{<}{_{\sim}}\else$\stackrel{<}{_{\sim}}$\fi}
\def\gapp{\ifmmode\stackrel{>}{_{\sim}}\else$\stackrel{>}{_{\sim}}$\fi}
\usepackage{graphicx}
\usepackage{multirow}
\usepackage{color}
\usepackage{amsmath}
\usepackage{soul}
\usepackage{hyperref}

\shorttitle{}
\shortauthors{}
\begin{document}

\title{X-ray studies of the pulsar PSR~J1420$-$6048 and its TeV pulsar wind nebula in the Kookaburra region}

\correspondingauthor{Hongjun An}
\email{hjan@cbnu.ac.kr}

\author{Jaegeun Park}
\author{Chanho Kim}
\affiliation{Department of Astronomy and Space Science, Chungbuk National University, Cheongju, 28644, Republic of Korea}
\author{Jooyun Woo}
\affiliation{Columbia Astrophysics Laboratory, Columbia University, New York, NY 10027, USA}
\author{Hongjun An}
\affiliation{Department of Astronomy and Space Science, Chungbuk National University, Cheongju, 28644, Republic of Korea}
\author{Kaya Mori}
\affiliation{Columbia Astrophysics Laboratory, Columbia University, New York, NY 10027, USA}
\author{Stephen P. Reynolds}
\affiliation{Physics Department, NC State University, Raleigh, NC 27695, USA}
\author{Samar Safi-Harb}
\affiliation{Department of Physics and Astronomy, University of Manitoba, Winnipeg, MB R3T 2N2, Canada}

\begin{abstract}
We present a detailed analysis of broadband X-ray observations of the pulsar PSR~J1420$-$6048 and its wind nebula (PWN) in the Kookaburra region
with Chandra, XMM-Newton, and NuSTAR. Using the archival XMM-Newton and new NuSTAR data,
we detected 68\,ms pulsations of the pulsar and characterized its X-ray pulse profile which exhibits a
sharp spike and a broad bump separated by $\sim$0.5 in phase.
A high-resolution Chandra image revealed a complex morphology of the PWN: a torus-jet structure, a few
knots around the torus, one long ($\sim$7$'$) and two short tails extending in the north-west direction,
and a bright diffuse emission region to the south.
Spatially integrated Chandra and NuSTAR spectra of the PWN out to 2.5$'$ are well described by a power law model with a photon index $\Gamma\approx 2$.
A spatially resolved spectroscopic study, as well as NuSTAR radial profiles of the 3--7\,keV and 7--20\,keV brightness, showed a hint of spectral softening with increasing distance from the pulsar.
A multi-wavelength spectral energy
distribution (SED) of the source was then obtained by supplementing our X-ray measurements with published radio, Fermi-LAT, and H.E.S.S. data.
The SED and radial variations of the X-ray spectrum
were fit with a leptonic multi-zone emission model. Our detailed study of the PWN
may be suggestive of (1) 
particle transport dominated by advection, (2)  
a low magnetic-field strength ($B \sim 5\mu$G), and (3) electron acceleration to $\sim$PeV energies.
\end{abstract}

\bigskip
\section{Introduction}
\label{sec:intro}
Ultra high-energy cosmic rays (UHECRs) with energies of $\gtrsim 10^{20}$\,eV
are detected on Earth, but their origin remains unclear.
It is well known that very high-energy cosmic-ray electrons ($\lapp10^{15}$\,eV) are produced
in pulsar-wind nebulae (PWNe) as evidenced by their TeV emission.
Acceleration of particles to very high energies in PWNe is thought to occur
at the termination shocks of their relativistic winds \citep[][]{kc84a}. The particles flow outwards
and form an extended bubble of synchrotron
radiation via interaction
with a magnetic field ($B$) 
as observed in the radio to X-ray band.
The particles can inverse-Compton upscatter (ICS) the synchrotron, CMB, and/or infrared (IR) photons to produce VHE emission \citep[e.g.,][]{Harding1996}.
This synchrotron-ICS emission scenario has been widely employed in models
of PWN emission and has provided useful information
on particle acceleration and transport in PWNe \citep[e.g.,][]{Gelfand2009,baa11}.

Observatories operating at 
ultra-high TeV energies \citep[e.g.,][]{Cao2021} have
revealed numerous TeV PWNe that can provide insights into cosmic PeVatrons, or
the origin of cosmic rays at energies of $\sim10^{15}$\,eV.
Hence, studies of PeV cosmic-ray electrons have been done primarily with observations
of very high-energy (VHE; $>$100\,GeV) radiation from these PWNe
and their surrounding halos \citep[e.g.,][]{HESS2018}.
In particular, very high-energy particles in PWNe with ages of $\sim$10--100\,kyr may escape
from the compact PWNe into the interstellar medium,
propagate towards Earth, and be detected as high-energy cosmic rays \citep[e.g.,][]{Giacinti2020}.

While VHE emission of PWNe is certainly a useful probe to explore particle
acceleration and propagation, properties of the particles at the highest energies cannot be precisely
characterized by VHE spectra alone because their emission spectra can be distorted by Klein-Nishina
suppression \citep{KN1929}. Besides, modeling the VHE emission 
requires knowledge of the ambient IR seed photon sources, which typically
have poorly known  temperature and density profiles. 
On the contrary, synchrotron emissions of PWNe do not suffer from the aforementioned effects and thus provide a complementary tool for investigating particle acceleration and transport in PWNe  \citep[e.g.,][]{Reynolds2016}.
The highest-energy particles (in TeV--PeV energies) emit synchrotron photons in the X-ray to MeV band, and hence
X-ray data are crucial to understanding UHECRs in PWNe \citep[e.g.,][]{Mori2021,Burgess2022}.

The point source PSR~J1420$-$6048 (J1420 hereafter)
and surrounding PWN-like emission were discovered by targeted X-ray observations of an EGRET source \citep[][]{Roberts1998,Roberts2001a}
in the so-called `Kookaburra' region \citep[][]{Roberts1999}. 
Radio pulsations with a period of 68\,ms were detected from J1420 \citep[][]{DAmico2001};
the radio pulsar, whose 
characteristic age ($\tau_c$) is $13$\,kyr,
is very energetic, with a high spin-down luminosity ($\dot E_{\rm SD}$)
of $10^{37}\rm \ erg\ s^{-1}$.
The discovery of the pulsations firmly established the association between J1420
and the nebula detected in the radio and X-ray bands \citep[the `K3' PWN;][]{Roberts1999,Roberts2001}.
\citet[][]{DAmico2001} estimated the distance to the pulsar to be
7.7\,kpc based on dispersion measure (DM), while 
5.6\,kpc was later suggested using a different DM model \citep[][]{Ng2005}.
Gamma-ray pulsations of J1420 were detected with high significance
\citep[][]{Weltevrede2010}\footnote{https://www.slac.stanford.edu/$\sim$kerrm/fermi\_pulsar\_timing/}
but an X-ray detection of the pulsations made with ASCA data was only marginal with
a chance probability of $p=0.0056$ \citep[][]{Roberts2001}. The X-ray pulsations
have not been confirmed by later studies \citep[][]{Ng2005,kh15}. These previous attempts to find X-ray pulsations seem to be hampered by the lack of photon statistics and/or strong contamination from the PWN.

Multi-band studies have been carried out to understand the K3 PWN.
\citet{Roberts1999} and \citet{VanEtten2010} measured radio flux densities of the source, and
its X-ray spectrum with a photon index $\Gamma\approx 2$ was seen to 
soften with increasing distance from the pulsar \citep[][]{VanEtten2010,Kishishita2012}.
\citet{hessrabbit2006} discovered an extended TeV source coinciding with K3
(but with an offset center) and measured a $\Gamma=2.2$ power-law spectrum in the TeV band.
\citet{VanEtten2010} modeled the broadband SEDs of
an inner region and an extended nebula of K3 using a two-zone
time-dependent emission model. They found that a leptonic scenario provides a better fit to the SEDs 
than a hadronic hybrid model.
In multi-band images, \citet{VanEtten2010} identified an apparent radio shell structure
and a long $8'$ X-ray tail in the north. These images suggest that the pulsar was born 3$'$ northwest of its current position
at the center of the apparent radio shell, and electrons spewed by the pulsar produce the X-ray tail and offset TeV emission via the synchrotron and ICS emission, respectively. 

In this paper, further X-ray investigations along with modeling multi-band SED and X-ray morphology data are attempted to understand the K3 PWN's properties better.
We describe data reduction processes
in Section~\ref{sec:sec2_1}, and present the data analyses and results
in Section~\ref{sec:sec2_2}--\ref{sec:sec2_4}.
We then apply a multizone model to the broadband SED and radial profiles, 
and infer the properties of the source (Section~\ref{sec:sec4}).
We discuss the results in Section~\ref{sec:sec5} and summarize in Section~\ref{sec:sec6}.
 
\section{X-ray Data Analysis}
\label{sec:sec2}

\subsection{Data reduction}
\label{sec:sec2_1}
We analyzed archival Chandra and XMM-Newton data acquired on 2010 December 8 for 90\,ks (Obs. ID 12545)
and on 2018 February 19 for 91\,ks (Obs. ID 0804250501), respectively,
and a new NuSTAR observation acquired on 2021 May 11 for 130\,ks \citep[Obs. ID 40660002002;][]{Mori2021}.
We reprocessed the Chandra data using {\tt chandra\_repro}
of CIAO~4.14 along with CALDB~4.9.6 for the most recent calibration data. 
The XMM-Newton data were processed with the {\tt emproc} and {\tt epproc}
tasks of SAS~20211130\_0941. Since the XMM-Newton observation  data were severely affected
by particle flares, we used tight flare cuts (e.g., {\tt RATE<=0.1}) which reduce the exposures substantially to $<$40\,ks. Still, the data suffer from some contamination by residual flares,
which could be a concern for spectral analysis of the faint and extended PWN. 
The flare background is less concerning for timing analysis of the pulsar, so
we perform timing analysis with the PN data after
applying a more typical flare cut of 
{\tt RATE<=0.4}.\footnote{https://www.cosmos.esa.int/web/xmm-newton/sas-thread-epic-filterbackground}
The NuSTAR data were reduced with {\tt nupipeline} in HEASOFT~v6.29
using the {\tt SAA\_MODE=strict} flag as recommended by the NuSTAR
science operation center. We verified that using {\tt SAA\_MODE=optimized}
did not change the results significantly (but see Section~\ref{sec:sec2_2}).
Net exposures after this reduction are
90\,ks, 65\,ks, and 57\,ks for Chandra, XMM-Newton/PN, and NuSTAR, respectively. 
Note that all errors reported in this paper are 1$\sigma$.

\subsection{Detection of the pulsations of J1420}
\label{sec:sec2_2}
\begin{figure}
\centering
\includegraphics[width=3.05 in]{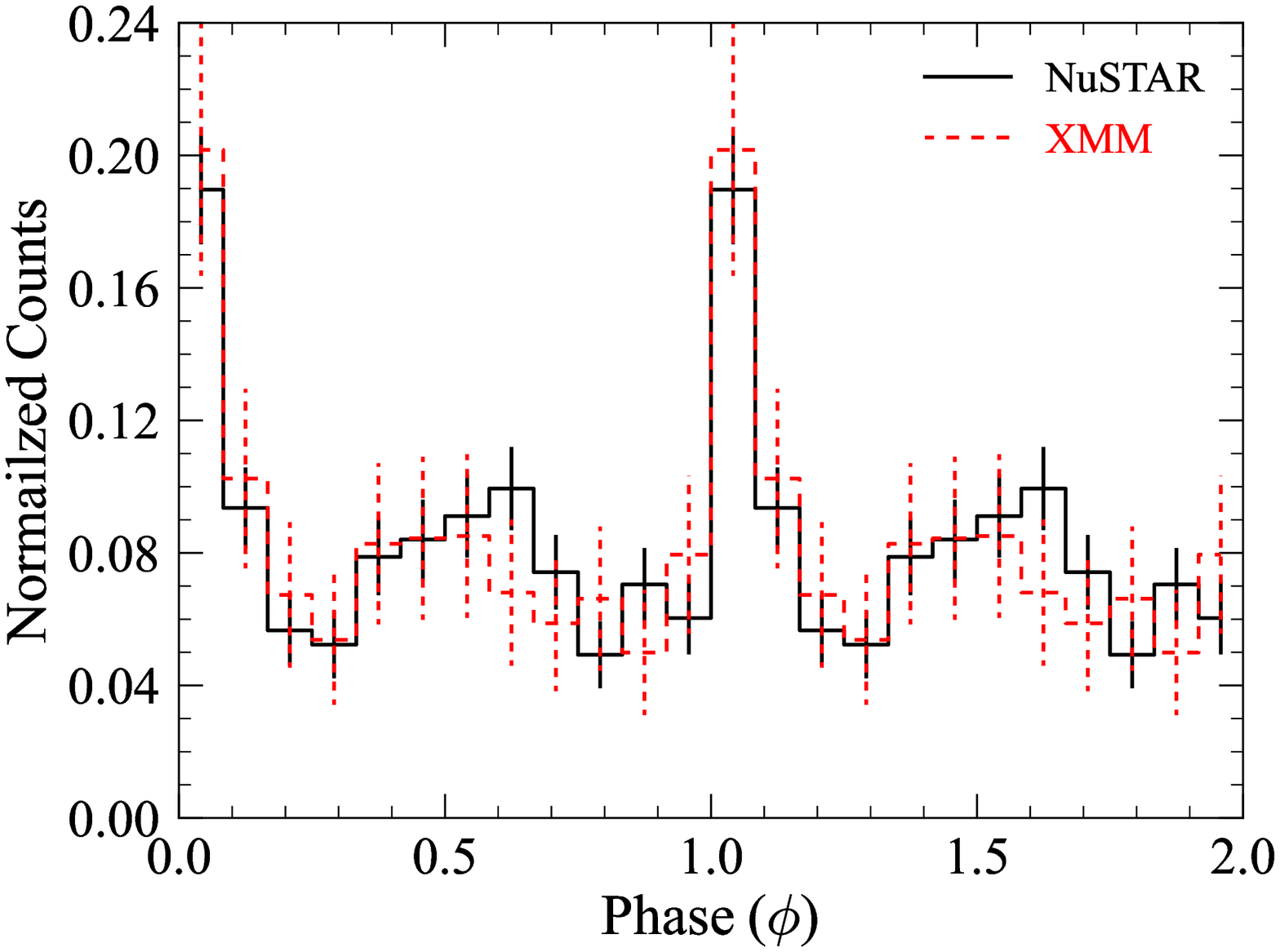} \\
\includegraphics[width=3.05 in]{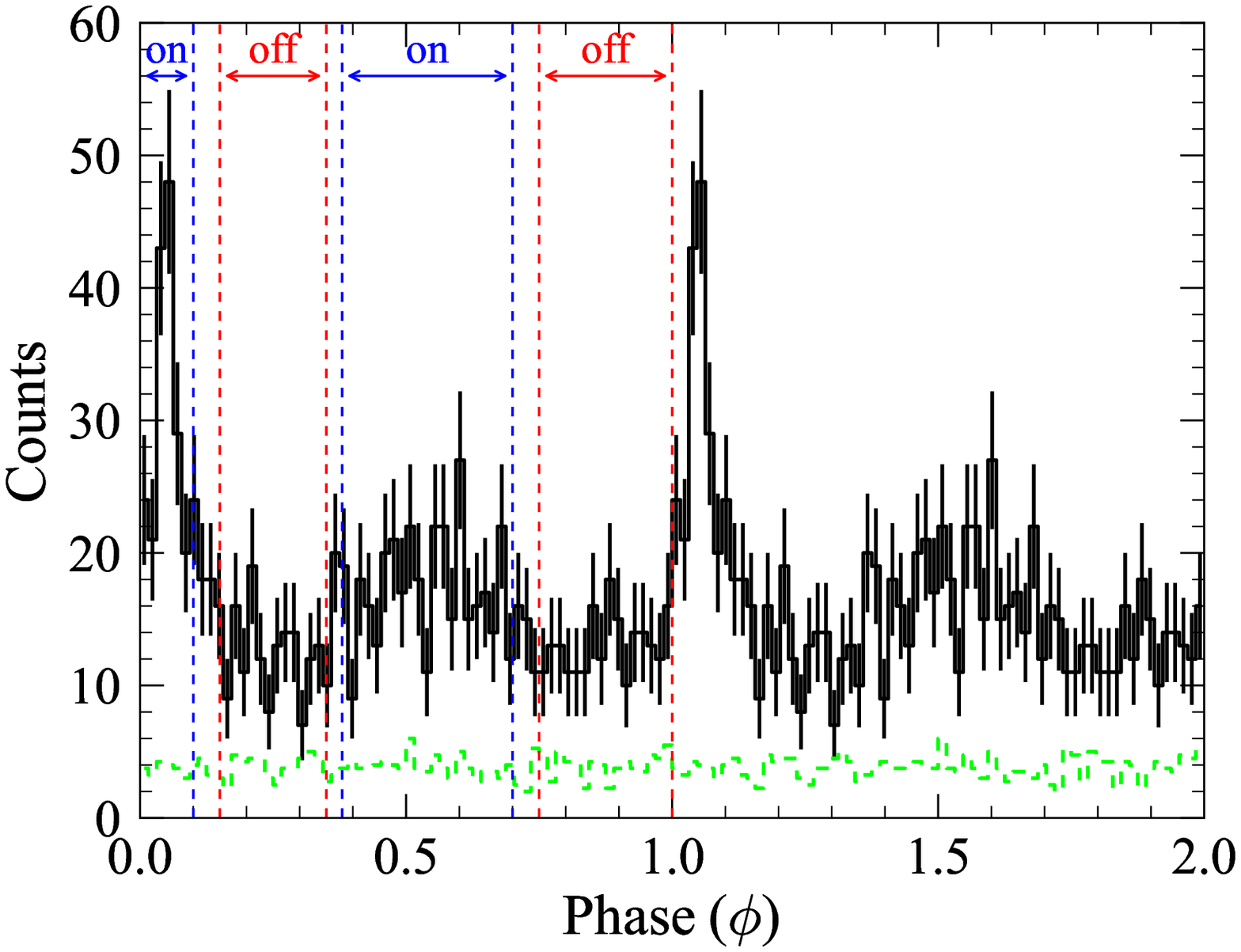} \\
\figcaption{{\it Top}:
1--10 keV XMM-Newton (red) and 3--30 keV NuSTAR (black; FPMA and FPMB combined) pulse profiles.
The XMM-Newton profile was constructed using photon weights (see text),
and the background was subtracted from the NuSTAR profile.
The profiles were normalized to have a summed count of 1, and reference phases
of the profiles were adjusted to align them.
{\it Bottom}: 3--30 keV NuSTAR profiles of the source (black) and background (green) events
measured using finer bins. \label{fig:fig1}}
\vspace{0mm}
\end{figure}

We searched the X-ray data for 68-ms pulsations of J1420 to
confirm an earlier X-ray detection which was only suggestive \citep[][]{Roberts2001}.
For the XMM-Newton data, we extracted 1--10\,keV events within a $R=16''$ circle centered
at the pulsar position of (R.A., decl.)=($215.034142^\circ$, $-60.804124^\circ$) \citep[][]{DAmico2001}
and applied a barycentric correction to the event arrival times.
We then folded the arrival times using a range of test frequencies ($f=14.656955$--$14.657095$\,Hz) around
one obtained by extrapolating the timing solution
\citep[][]{krjs+15} obtained by the Fermi Large Area Telescope \citep[LAT;][]{fermimission}, 
and computed an $H$ statistic \citep[][]{drs89} for each test frequency. In this search, we
held the frequency derivative $\dot f$ fixed at the LAT-measured value of $\dot f = -1.772\times 10^{-11} \rm \ Hz\ s^{-1}$.
This search resulted in a significant detection at $f=14.657082$\,Hz (MJD~58168)  with $H\approx40$, corresponding to a post-trial chance probability of $p\approx5\times 10^{-7}$.
The detection significance changes substantially between $H\approx30$ and $H\approx60$
depending on the region selection, presumably because of strong PWN emission.

In order to mitigate contamination by the PWN,
we adopted a weighted $H$ test \citep{k11}, for which the probability of each event being
a source photon is computed by fitting an image of a region around J1420 with the point spread function (PSF)
plus a constant.
This yielded more stable (reliable) detection of
the pulsations with $H\approx50$--$60$, corresponding to a post-trial $p\approx10^{-9}$,
regardless of the region selection as long as the region size was reasonable (e.g., $R\ge 10''$).
The probability-weighted 1--10\,keV pulse profile is displayed in Figure~\ref{fig:fig1} (top).

We expected that X-ray pulsations might be more easily detected in the NuSTAR data since
the pulsar's spectrum was inferred to be hard \citep[$\Gamma\approx0.5$;][]{kh15}.
In NuSTAR images (Section~\ref{sec:sec2_3_2}), excess emission at the pulsar position was noticeable up to $\sim$30\,keV whereas the extended PWN was not significantly detected at energies above 
20\,keV, meaning that the pulsar emission is spectrally harder than the PWN emission. We therefore used the 3--30\,keV and 3--20\,keV bands for the pulsar and the PWN emission (e.g., Section~\ref{sec:sec2_4_2}), respectively.
We extracted 3--30\,keV source events
within a $R=30''$ circle centered at the brightest spot in the 10--30\,keV smoothed images,
and performed an $H$ test in a range of $f$ (14.655187--14.655320\,Hz)
holding $\dot f$ fixed at the LAT-measured value.
Source pulsations with $f=14.655289\rm \ Hz$ on MJD~59345 were more significantly
detected in the NuSTAR data ($H\approx 100$ corresponding to a post-trial $p=6\times 10^{-17}$)
than in the XMM-Newton data. The XMM-Newton and NuSTAR measurements of
$f$ yielded an average frequency derivative of $-1.763\times 10^{-11}\rm \ Hz\ s^{-1}$
which is similar to that measured by Fermi-LAT.
The 3--30\,keV pulse profile is displayed in Figure~\ref{fig:fig1}.
We also checked to see if the pulsations persist at $>$20\,keV.
While the 20--30\,keV pulse profile seemed to show a hint of the pulsations,
the detection significance was low ($H\le$10), likely due to the paucity of counts.
We therefore relaxed the {\tt SAA\_MODE} cut (from {\tt strict} to {\tt optimized}\footnote{https://heasarc.gsfc.nasa.gov/docs/nustar/analysis/nustar\_sw\\guide.pdf}) and were able to
detect the 20--30\,keV pulsations significantly with
$H\approx 20$ ($p\approx 3\times 10^{-4}$).

We arbitrarily adjusted the reference phases of the XMM-Newton and NuSTAR
profiles to align their peaks, and display them in Figure~\ref{fig:fig1}.
The profiles show a sharp peak at phase $\phi\approx0.05$ and a broad bump (Fig.~\ref{fig:fig1}).
The bump is visible in both the XMM-Newton and NuSTAR data,
but a constant function can fit the profiles in the phase interval for the bump ($\phi=0.25$--$0.9$)
with $p$=0.01 and 0.9 for the NuSTAR and XMM-Newton profile, respectively; the existence of
the broad bump is not definitive.
A further investigation of the profile with the high timing resolution NuSTAR data
revealed that the sharp peak is indeed narrow ($\Delta \phi\lapp0.1$; Fig.~\ref{fig:fig1} bottom). 

\begin{figure*}
\centering
\hspace{-3.0 mm}
\includegraphics[width=7.0 in]{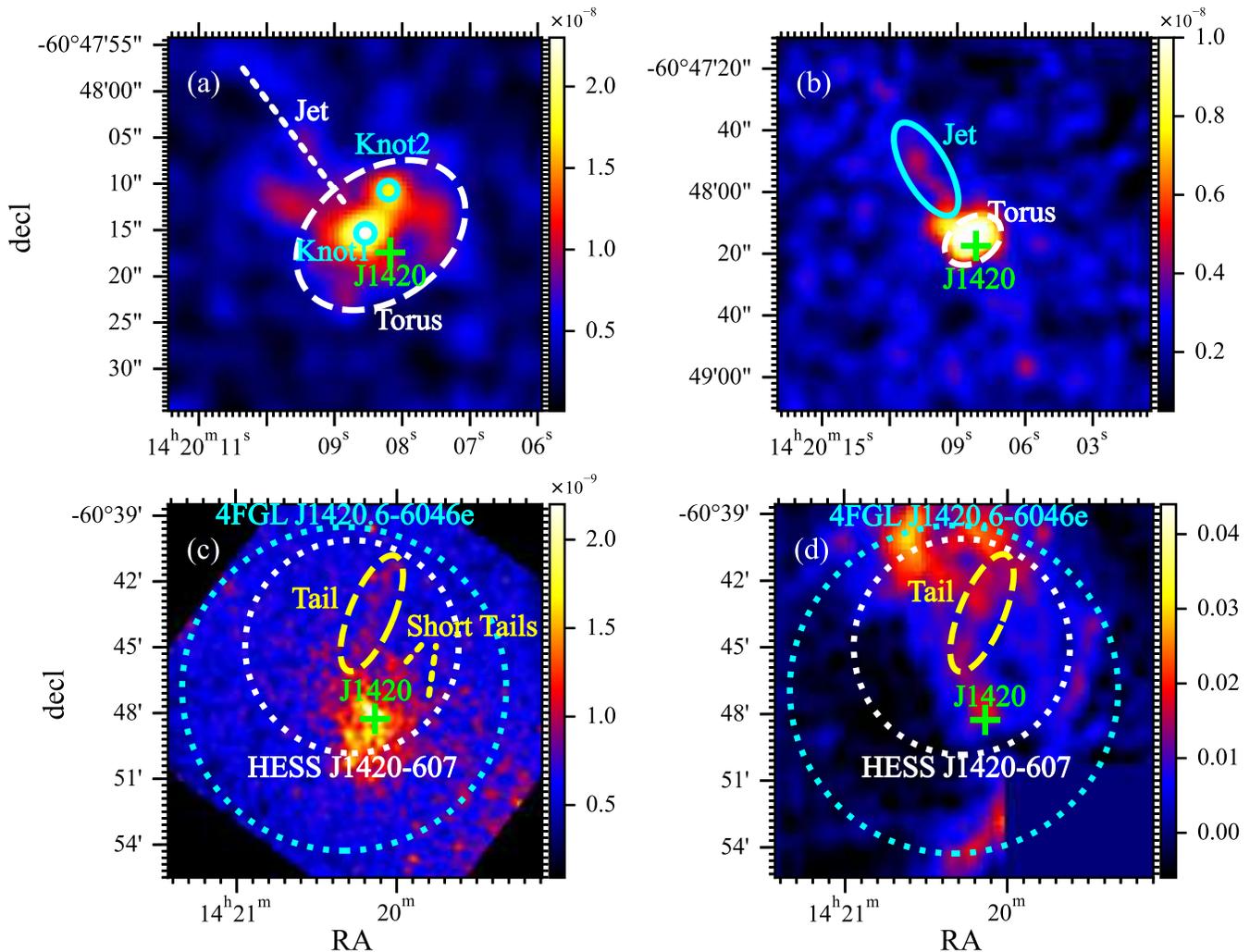}
\figcaption{2--7\,keV Chandra images of the K3 PWN on various spatial scales (a--c) and a SUMSS 843\,MHz image (d).
The images are smoothed and the scales are adjusted for better legibility.
The point-source-removed and exposure-corrected Chandra images are constructed following a CIAO science thread.
The position of J1420 is denoted by a green cross.
({\it a}) Chandra image on a $40''\times 40''$ scale.
In addition to the two knots denoted in cyan, a faint jet (not visible in this panel) and a torus are marked. 
({\it b}) Chandra image on a $2'\times2'$ scale. The torus and the jet structures
are shown in white and cyan ellipses, respectively.
({\it c}) Chandra image on a $16'\times 16'$ scale.
A long northern tail region is denoted by a yellow ellipse and two short tails are marked.
Regions of the Fermi-LAT ($0.123^\circ$; radius of a disk model)
and H.E.S.S. ($0.08^\circ$; 1$\sigma$ width of a Gaussian model) counterparts are displayed in cyan and white circles, respectively.
({\it d}) SUMSS 843\,MHz image \citep[][]{Mauch2003}. The image was downloaded from the Skyview webpage and truncated to match the X-ray emission region in panel (c).
Note also that a box region in the lower right corner is excised because it contains a bright unrelated point source.
The pulsar, northern X-ray tail, and Fermi-LAT and H.E.S.S. regions are overlaid for comparison.
\label{fig:fig2}
}
\vspace{0mm}
\end{figure*}

\subsection{Image analysis}
\label{sec:sec2_3}
Previous Suzaku studies of K3 \citep[][]{VanEtten2010,Kishishita2012} found that the source
is elongated to the north where an apparent radio shell lies \citep[][]{VanEtten2010}.
On a smaller scale, \citet{Ng2005} found an X-ray arc in 10\,ks Chandra data (Obs. ID 2792) and
suggested that it could be the termination shock.
Moreover, the spectral softening detected in K3 \citep[][]{VanEtten2010,Kishishita2012} may
be manifested by a size shrinkage with increasing energy \citep[e.g.,][]{nhra+14,amrk+14}.
In this section, we inspect the Chandra data to identify the small- and large-scale structures,
and analyze the NuSTAR data to measure the size shrinkage of K3.

\subsubsection{Chandra images}
\label{sec:sec2_3_1}
We produced a
Chandra image in the 2--7\,keV band after removing point sources
and correcting for exposure. This was done by following
a procedure in the CIAO science thread.\footnote{https://cxc.cfa.harvard.edu/ciao/threads/diffuse\_emission/}
We then adjusted the image scales and bins to identify structures in the PWN
on various spatial scales (Fig.~\ref{fig:fig2}).

On a $40''\times40''$ scale, we identified two knots at 3$''$ and 7$''$ from the pulsar
(Knot 1 and Knot 2; Fig.~\ref{fig:fig2} a); compared to nearby backgrounds, the knots
were detected at $\sim$3$\sigma$ significance.
Note that Knot 1 is listed
in the Chandra source catalog\footnote{https://cxc.cfa.harvard.edu/csc/},
but we do not find an IR or optical counterpart in the 2MASS and USNO catalogs.
Furthermore, the knots appear to be more extended than nearby point sources with similar counts.
Note also that there seems to be a slightly elongated structure (short and parallel to but just east of the jet indicated in Figure~\ref{fig:fig2}b), but this structure was detected only at the 2.5$\sigma$ level.
An image of a $2'\times2'$ region near the pulsar is displayed in Figure~\ref{fig:fig2}b. In this image,
the torus structure in the east-west direction identified by \citet{Ng2005} is clearly visible.
In addition, we found a narrow jet-like ($\sim20''$) feature extending in the north-east direction.
The jet-like structure is fainter but detected at a $\ge3\sigma$ level,
having 172 events in the 1--10\,keV band 
within a $7''\times 17''$ ellipse (excluding the pulsar and torus emission; Fig.~\ref{fig:fig2} b)
which contains estimated 119 background counts.

A larger-scale image is displayed along with VHE
emission regions \citep[][]{HESS2018,fermi4fgl} in Figure~\ref{fig:fig2} c.
The image reveals a prominent $\sim$7$'$ tail (denoted as `Tail')
and two short tails in the north-west direction.
These tails seem to constitute the broad northern tail observed in the
low-resolution Suzaku image \citep[][]{VanEtten2010, Kishishita2012}.
In the south, a bright emission region is seen out to $\sim2.5'$.
The bright southern region and the northern `Tail' were also identified in our inspection of XMM-Newton MOS images. The `Tail' appears to partially overlap with a radio structure \citep[Fig.~\ref{fig:fig2} d; see also][]{VanEtten2010}.\footnote{see https://skyview.gsfc.nasa.gov/current/cgi/morein\\fo.pl?survey=SUMSS\%20843\%20MHz for more information on the radio image}
The VHE emission regions are centered
at 2--3$'$ north of the pulsar (white and cyan circles in Fig.~\ref{fig:fig2} c) and also overlap well
with the X-ray PWN.
Note also that there is weak excess emission in the southwest (near the chip boundary). This region is 15--20\% brighter than other background regions, possibly indicating inhomogeneous sky emission on a larger scale as was seen in an ASCA image \citep[e.g.,][]{Roberts2001a}.

We also searched the Chandra images for a shell-like structure (e.g., supernova remnant) in several energy bands
but did not find any. Additionally, we compared spectra of various regions within the FoV with blank-sky data\footnote{https://cxc.cfa.harvard.edu/ciao/threads/acisbackground/} to see if there is an excess of line emission but found none.

\subsubsection{NuSTAR image}
\label{sec:sec2_3_2}
\begin{figure}
\centering
\hspace{-3.0 mm}
\includegraphics[width=3.1 in]{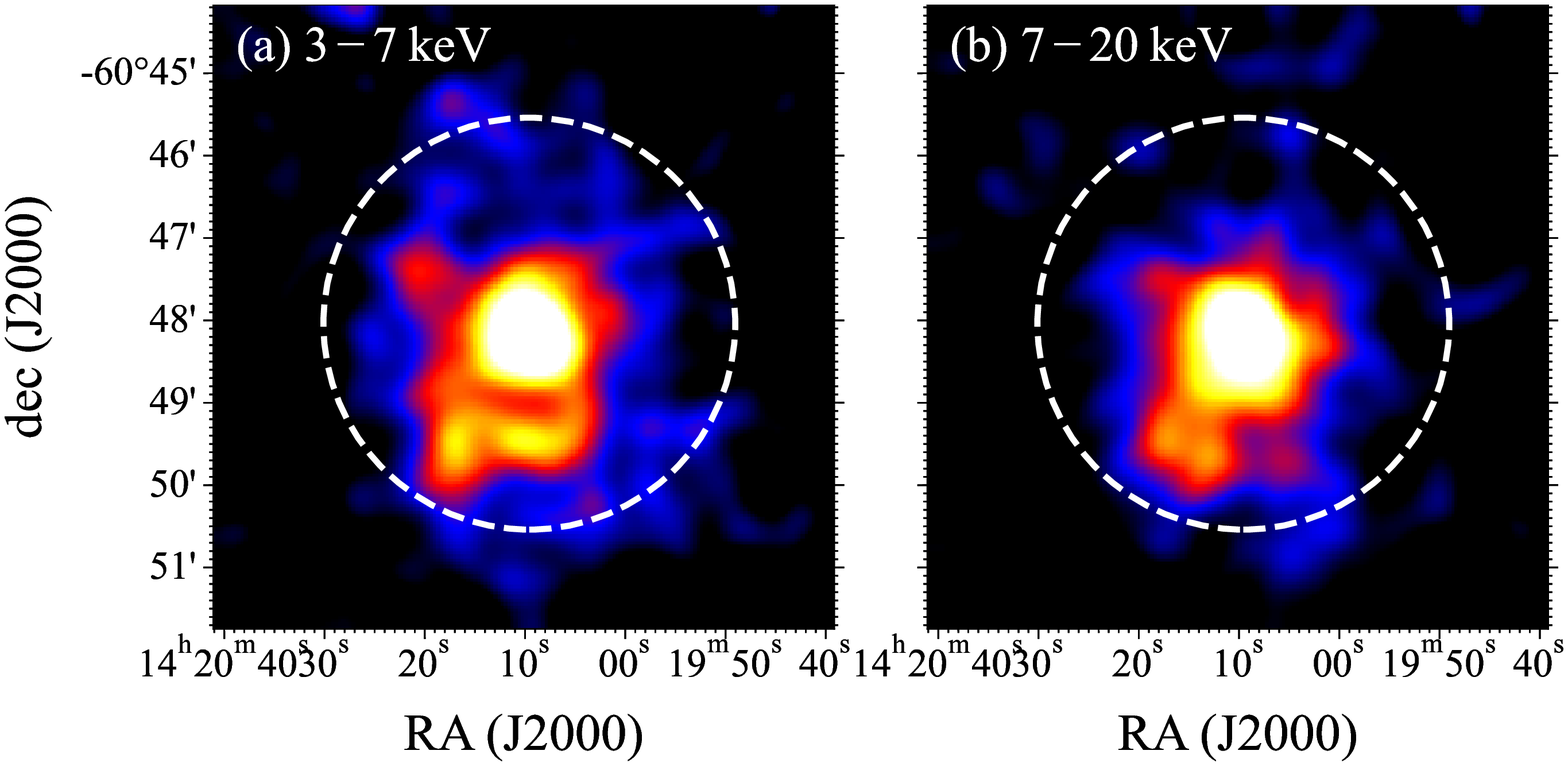} 
\includegraphics[width=3.1 in]{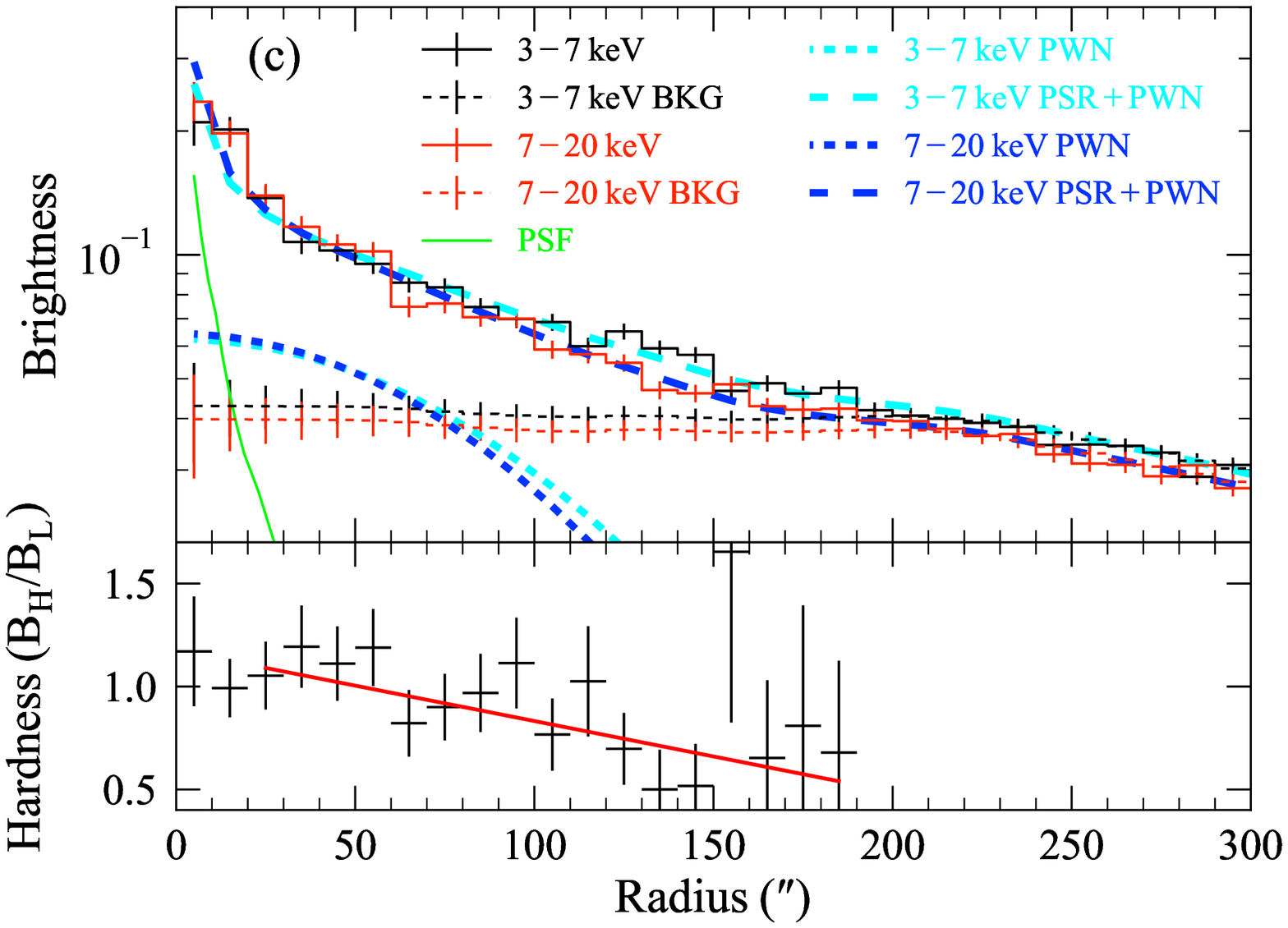}
\figcaption{Background-subtracted NuSTAR images (a--b)
and radial profiles (c) in the 3--7\,keV and 7--20\,keV bands.
({\it a}--{\it b}) NuSTAR images in the 3--7\,keV (a)
and 7--20\,keV bands (b).
The images are smoothed and normalized to 1 at the maximum counts, and $R=2.5'$ circles
are shown for reference. The image scales are adjusted for legibility.
({\it c}) 3--7\,keV (black) and 7--20\,keV (red) radial profiles of
surface brightness (counts per area) are presented in the top panel.
The green line is the radial profile of the PSF (pulsar), and
the black and red dotted lines are background profiles
in the low and high-energy bands, respectively.
Radial Gaussian functions that describe the PWN profiles in the low- and high-energy bands
are shown in cyan and blue dotted lines, respectively. The summed models are presented in the cyan and blue dashed lines for the low- and high-energy profiles, respectively. 
Hardness ratios, defined by the high-energy to low-energy brightness
ratio, are displayed in the bottom panel, and the red line is
a linear fit to the hardness ratios.
\label{fig:fig3}}
\vspace{-2mm}
\end{figure}

We next produced 3--7\,keV and 7--20\,keV NuSTAR images of the source.
To take into account the spatially non-uniform background, we carried out {\tt nuskybgd}
simulations.\footnote{https://github.com/NuSTAR/nuskybgd}
Although the simulated background images reproduced the aperture patterns well, the normalization of the simulated image differed from the observed one 
by $<$10\% in each chip.
Hence, we manually adjusted the background normalization to match
the observed background counts in each chip.
We combined FPMA and FPMB images after aligning them using the brightest spot
in the 10--30\,keV smoothed images (Section~\ref{sec:sec2_2}).
Background-subtracted NuSTAR images of the K3 PWN show primarily
the bright southern region (Fig.~\ref{fig:fig3}), and
the low- and high-energy morphologies do not appear significantly different.

We produced 3--7\,keV and 7--20\,keV radial profiles
to further investigate possible size shrinkage of K3 with energy.
Although contamination from the pulsar can be minimized
by pulse gating, the statistics would then be insufficient
for a meaningful comparison because the off-pulse intervals are narrow
($\Delta \phi < 0.5$; Fig.~\ref{fig:fig1} top).
Therefore, we used all the pulse phases.
Radial profiles in the low- and high-energy bands along with
simulated background profiles are displayed in the top panel of Figure~\ref{fig:fig3} c;
the background profiles account for the observations at large radii as expected.
To assess the relative contributions of the pulsar and extended emission, we fit the radial profile with a model composed of
the PSF (pulsar), background, and a Gaussian function $A\mathrm{exp}(-r^2/2w^2)$ (PWN);
the best-fit functions are displayed in Figure~\ref{fig:fig3} c.
The inferred normalization factors for the background profiles were consistent with 1
at $\lapp 1\sigma$ level, and the pulsar flux was estimated to be $\approx$10\% of the PWN flux (see also Table~\ref{ta:ta1}) and dominates only in the innermost regions $r<20''$ (Fig.~\ref{fig:fig3} c).

The widths ($w$) of the best-fit Gaussian functions for the low- and high-energy profiles
were determined to be $82\pm3''$ and $75\pm2''$, respectively.
The measured widths differ only at the 1.6$\sigma$ level and thus do not clearly require a size shrinkage with energy. Because the synchrotron burn-off effects should also produce spectral softening with increasing radius, we computed hardness ratios (ratio of the hard- and soft-band brightness)
and show them in the bottom panel of Figure~\ref{fig:fig3} c. The hardness ratio nearly monotonically
decreases out to $\sim$200$''$ beyond which the background dominates.
A linear fit to the hardness ratios ($20''\le r\le 190''$) found a negative slope (i.e., spectral softening)
at the $\approx 3\sigma$ level with $\chi^2$/dof=10/17 (red line in the bottom panel of Fig.~\ref{fig:fig3} c).
Note, however, that the negative slope might be caused primarily by the two data points at $R=130''$--$140''$,
rather than by a gradual decrease. Ignoring the two points reduced the significance to a $\approx 2\sigma$ level,
which still hints at a gradual decrease although a firm conclusion on the softening cannot be made with the current data.

\subsection{Spectral analysis}
\label{sec:sec2_4}
In this section, we analyze X-ray spectra of the K3 PWN and
its sub-structures that were identified in the Chandra images (Fig.~\ref{fig:fig2}).
The spectral softening measured for K3 \citep[][]{VanEtten2010, Kishishita2012} implies a spectral curvature in its spatially integrated spectrum which may be detected in the broadband X-ray data taken with NuSTAR \citep[e.g.,][]{mrha+15}. Furthermore, the NuSTAR data may allow the detection of a spectral cut-off at $>$10\,keV \citep[e.g.,][]{a19}.
Because other point sources within the PWN may contaminate the NuSTAR spectra of the PWN, we estimate the contributions from the point sources 
using the high-resolution Chandra data.

\newcommand{\marka}{\tablenotemark{a}}
\newcommand{\markb}{\tablenotemark{b}}
\newcommand{\markc}{\tablenotemark{c}}
\newcommand{\markd}{\tablenotemark{d}}
\newcommand{\marke}{\tablenotemark{e}}
\newcommand{\markf}{\tablenotemark{f}}
\begin{table*}[t]
\vspace{-0.0in}
\begin{center}
\caption{Spectral analysis results}
\label{ta:ta1}
\vspace{-0.05in}
\scriptsize{
\begin{tabular}{lcccccc} \hline\hline
data       & Instrument\marka  & energy range   & $N_{\rm H}$      & $\Gamma$  &  $F_{\rm 3-10\,keV}$ & $\chi^2$/dof \\
           &        & (keV)   & ($10^{22}\rm \ cm^{-2}$) &  & ($10^{-13}\rm \ erg\ s^{-1}\ cm^{-2}$) &  \\ \hline
PSR\markb  & C      & 0.5--10 & $4.6$\markc & $0.7\pm0.2$   & $1.3^{+0.2}_{-0.1}$   & 30/46 \\
PSR\markd  & N      & 3--30   & $4.6$\markc & $0.7\pm0.4$   & $1.5^{+0.6}_{-0.4}$   & 21/22 \\ \hline
PWN\marke  & C      & 0.5--10 & $4.2\pm0.5\pm0.4$ & $1.82\pm0.23\pm0.11$ & $15.3\pm0.9\pm0.4$   &  156/146  \\ 
PWN\marke        & N      & 3--20   & $4.2$\markf  & $1.97\pm0.06^{+0.08}_{-0.07}$ & $16.5\pm0.6\pm0.6$  & 168/170  \\ 
PWN\marke  & C+N    & 0.5--20 & $4.6\pm 0.3\pm0.4$ & $1.98\pm0.07^{+0.08}_{-0.06}$  & $15.0\pm0.7\pm0.3$ & 324/317 \\ \hline
Torus      & C      & 0.5--10 & $4.6$\markc & $1.8\pm0.3$   & $0.7\pm0.1$   & 6/10  \\ 
Jet        & C      & 0.5--10 & $4.6$\markc & $2.1\pm0.7$   & $0.2\pm0.1$ & 4/5   \\ 
Tail       & C      & 0.5--10 & $4.6$\markc & $2.06\pm0.26$  & $3.5^{+0.5}_{-0.4}$ & 134/120  \\  \hline
\end{tabular}}
\end{center}
\vspace{-0.5 mm}
\footnotesize{
$^{\rm a}${C: Chandra, N: NuSTAR.}\\
$^{\rm b}${On+off pulse emission.}\\
$^{\rm c}${Fixed at the value obtained from a joint fit of the Chandra+NuSTAR PWN spectra.}\\
$^{\rm d}${On$-$off pulse emission.}\\
$^{\rm e}${The second error is a systematic uncertainty. See text for more detail.}\\
$^{\rm f}${Fixed at the Chandra-measured value.}\\}
\end{table*}

\subsubsection{Point sources within the PWN}
\label{sec:sec2_4_1}
In the image analysis (Section~\ref{sec:sec2_3_1}), we 
detected 10 point sources and the pulsar J1420 within a $R=2.5'$ circle centered at the pulsar, using the {\tt wavdetect} tool of CIAO.
The point sources seen in the Chandra image are very faint, having 10--20 events within $R=2''$ regions compared to
260 events for J1420.
With so few counts, accurate spectral characterization of each source
was unfeasible. However, these faint sources are unlikely to affect  NuSTAR measurements of the PWN spectra
(Sections~\ref{sec:sec2_4_2} and \ref{sec:sec2_4_4}).
For a better assessment, we stacked the Chandra spectra 
of the 10 point sources using $R=2''$ extraction regions.
A summed background spectrum was constructed using $R=3''$ circles near the
source regions. We grouped the stacked spectrum to have at least 5 counts per bin and fit the spectrum with an absorbed power-law model employing the $l$ statistic \citep[][]{l92}.
For the Galactic absorption, we adopted 
the {\tt tbabs} model along with the {\tt vern} cross section \citep[][]{vfky96}
and {\tt angr} abundances \citep[][]{angr89} in XSPEC~v12.12.0 (throughout this paper).
Although the hydrogen column density $N_{\rm H}$ was not well constrained, the model fit favored a very
low value (consistent with 0) perhaps because of low-energy emission from a few soft, foreground sources -- these soft X-ray sources do not have a significant contribution to $>$3\,keV NuSTAR spectra. 
We, therefore, set $N_{\rm H}$ to 0 and found that a power law model with a photon index
of $1.2\pm 0.2$ and absorption-corrected 3--10\,keV flux
$F_{\rm 3-10\,keV}=2.3^{+0.6}_{-0.5}\times 10^{-14}\rm \ erg\ cm^{-2}\ s^{-1}$ fits the data;
the latter is $\le$2\% of the PWN flux (see below).
Note that using a larger $N_{\rm H}$ makes $\Gamma$ softer (i.e. fewer counts in the NuSTAR band)
and thus our estimation above is conservative.  

To measure the pulsar spectrum with the Chandra data, we extracted events within a $R=2''$ circle centered at
the pulsar position. A background spectrum was extracted within two $R=2''$ circles
in the torus region. We grouped the source spectrum to have a minimum of 5 counts 
per spectral bin and employed the $l$ statistic in the fit. The spectrum was well fit
with an absorbed power-law model for a frozen $N_{\rm H}=4.6\times 10^{22}\rm \ cm^{-2}$ which
was obtained from a joint fit of the Chandra+NuSTAR PWN spectra of a large region (see Section~\ref{sec:sec2_4_2}).
The fit resulted in $\Gamma=0.7\pm0.2$ and $F_{\rm 3-10\,keV}=1.3 \times 10^{-13}\rm \ erg\ cm^{-2}\ s^{-1}$
(Fig.~\ref{fig:fig4}).
Note that \citet{kh15}, using a likelihood method applied to the same Chandra data,
obtained $\Gamma=0.46\pm0.07$
for $N_{\rm H}=3.35^{+0.74}_{-0.51}\times 10^{22}\rm \ cm^{-2}$.
The difference in the measured photon indices seems to arise from its covariance with $N_{\rm H}$; by fixing $N_{\rm H}$ to  $3.35\times 10^{22}\rm \ cm^{-2}$, the photon index was fit to $\Gamma=0.4\pm0.2$, consistent with the results in \citet{kh15}. 

We also measured the pulsed spectrum of J1420 using the NuSTAR data.
We extracted source events within a $R=30''$ circle around the pulsar position and selected on- ($\phi$=0--0.1 and 0.38--0.7; Fig.~\ref{fig:fig1}) and off-pulse ($\phi$=0.15--0.35 and 0.75--1.00) data
for the source and background spectrum, respectively.
The pulsed (on$-$off) spectrum was fit with an absorbed power-law model in the 3--30\,keV band,
and we found the best-fit parameters to be $\Gamma=0.7\pm0.4$ and
$F_{\rm 3-10\,keV}=1.5^{+0.6}_{-0.4}\times 10^{-13}\rm \ erg\ cm^{-2}\ s^{-1}$.
The latter becomes $F_{\rm 3-10\,keV}=6^{+3}_{-2}\times 10^{-14}\rm \ erg\ cm^{-2}\ s^{-1}$
when averaged over a spin cycle (Fig.~\ref{fig:fig4}).

\subsubsection{Spatially integrated PWN spectrum}
\label{sec:sec2_4_2}
To construct a broadband X-ray SED to be used in our SED modeling (Section~\ref{sec:sec4}),
we measured the spatially integrated spectra of the PWN with the Chandra and NuSTAR data.
For the Chandra analysis, we extracted the source spectrum within a $R=2.5'$ circle
centered at the pulsar, excising the pulsar and other point sources (Section~\ref{sec:sec2_4_1}) using $R=2''$ circles.
This PWN region contains most of the bright nebula, but the outer part of
the long `Tail' is not included (see Section~\ref{sec:sec2_4_3} for the tail spectrum).
Because background selection for the faint and extended nebula was a concern,
we used the blank-sky data to choose optimal background regions by comparing
the blank-sky background and the observed image.
We note that the source region lies across four detector chips,
and hence we adjusted the background region sizes taken in the four chips
so that the sizes are approximately proportional to the source-region areas
within the corresponding chips. We verified that the blank-sky data explained well the instrumental line emissions \citep[e.g.,][]{Bartalucci2014} in the source and background regions and that the source and background spectra (including instrumental lines)
agreed very well in the low- ($\lapp$1.5\,keV) and high-energy ($\gapp$7\,keV) bands, meaning that
the background represents well the detector and sky background in the source region.

\begin{figure}
\centering
\includegraphics[width=3.3 in]{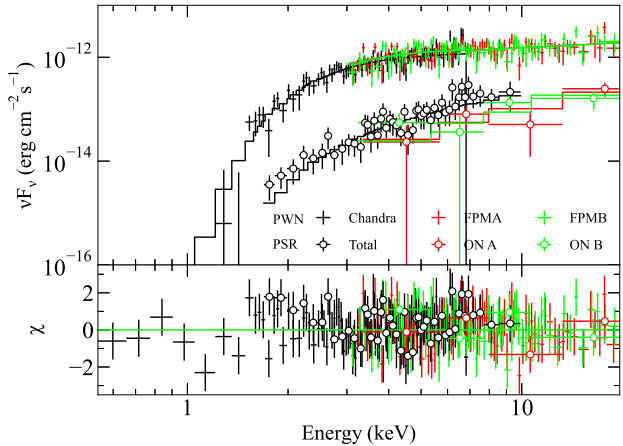}
\figcaption{Chandra (black) and NuSTAR (red and green) spectra of J1420 (empty circles) and the K3 PWN (crosses).
The NuSTAR spectra of the pulsar were generated in the on-pulse phases by subtracting the off-pulse background and were averaged over the spin cycle.
Best-fit models are presented as solid lines.
\label{fig:fig4}}
\vspace{0mm}
\end{figure}

We generated response files for the extended source region using the {\tt specextract} tool
of CIAO, grouped the spectrum to have a minimum of 100 counts per spectral bin, and
fit the 0.5--10\,keV spectrum with an absorbed power-law model (Fig.~\ref{fig:fig4}).
This model
adequately describes the observed spectrum ($\chi^2$/dof=156/146),
and the fit-inferred parameter values are $N_{\rm H}=(4.2\pm 0.5)\times 10^{22}\rm \ cm^{-2}$,
$\Gamma=1.82\pm 0.23$, and $F_{\rm 3-10\,keV}=(1.53\pm0.09)\times 10^{-12}\rm \ erg\ cm^{-2}\ s^{-1}$.
These $N_{\rm H}$ and $\Gamma$ values are similar to the previous measurements of
\citet{VanEtten2010} and \citet{Kishishita2012}.
Note, however, that they did not report the Galactic abundance and scattering cross-section used for their absorption models, and thus we assumed that
they used the {\tt vern} cross-section and {\tt angr} abundances (the default in XSPEC).
The PWN spectrum is better constrained by combining it with the NuSTAR data below.

As noted above, the Chandra-only fit results may significantly vary depending on the background selection,
especially because of the covariance between $N_{\rm H}$ and $\Gamma$.
Therefore, we analyzed the data with 10 different background selections to estimate
systematic uncertainties (i.e., 1$\sigma$ variation)
and found that the best-fit parameters change by $\Delta N_{\rm H}=\pm4\times 10^{21}\rm \ cm^{-2}$,
$\Delta \Gamma=\pm0.11$, and $\Delta F_{\rm 3-10\,keV}=\pm 4\times 10^{-14}\rm \ erg\ cm^{-2}\ s^{-1}$.
We report these systematic uncertainties as additional errors in Table~\ref{ta:ta1}.

To measure the NuSTAR spectrum of the PWN,
we used $R=2.5'$ circles for extraction of the source spectra (Fig.~\ref{fig:fig4}).
As noted above, the NuSTAR background is non-uniform,
and thus it was difficult to extract local background spectra 
on the same detector chip, especially for FPMA. We therefore used the
{\tt nuskybgd} simulations \citep[][]{Wik2014}
to generate a background spectrum corresponding to the source region
for each of FPMA and FPMB.
We verified that the background model adequately explained the observed background spectra
as well as the high-energy background ($>$30\,keV) in the source spectra.

Because the pulsar emission was included in our phase-integrated NuSTAR spectra,
we modeled those spectra with two power laws, one for the PWN and the other for the pulsar emission. We
held the second power-law parameters fixed at Chandra-measured pulsar parameters and fit the 3--20\,keV source spectra holding $N_{\rm H}$ fixed at $4.2\times 10^{22}\rm \ cm^{-2}$.
An acceptable fit ($\chi^2$/dof=168/170) was achieved with the best-fit parameters of
$\Gamma=1.97\pm0.06$ and $F_{\rm 3-10\,keV}=(1.65\pm0.06)\times 10^{-12}\rm \ erg\ cm^{-2}\ s^{-1}$.
The NuSTAR-measured flux and photon index are consistent with
the Chandra-measured ones
at the $\sim$1$\sigma$ level.
To check this further, we inspected the NuSTAR spectra in the 3--10\, keV band
and found that the fit-inferred PWN $\Gamma=1.92\pm0.10$ is still consistent with
the Chandra measurement.
We also attempted to fit the 3--20\,keV PWN spectra with a broken power-law model 
and found that a spectral break is not statistically required  with an $f$-test probability of $\approx 0.5$.

We assessed systematic uncertainties on the NuSTAR-inferred parameters.
The NuSTAR fit results are not very sensitive to a modest change of $N_{\rm H}$;
varying it within the Chandra-estimated uncertainty of $\pm7\times 10^{21}\rm \ cm^{-2}$ (Table~\ref{ta:ta1})
changes $\Gamma$ by $\pm0.04$
and $F_{\rm 3-10\,keV}$ by $\pm 5\times 10^{-14}\rm \ erg\ cm^{-2}\ s^{-1}$ (min/max).
The assumed pulsar spectral model may also introduce some uncertainties in the
inferred spectral parameters. We therefore varied the pulsar model
within the measurement uncertainties considering the covariance
between $\Gamma$ and $F_{\rm 3-10\,keV}$ of the pulsar.
We found that the effects of the pulsar model are not significant;
$\Gamma$ varies by $+0.05$ and $-0.03$, and the change of $F_{\rm 3-10\,keV}$ is
$\Delta F_{\rm 3-10\,keV}=\pm 10^{-14}\rm \ erg\ cm^{-2}\ s^{-1}$ (min/max).
We also varied background regions for the {\tt nuskybgd}
simulations \citep[see][for more detail]{Wik2014},
generated 20 background spectra, and used them in our spectral fits.
The best-fit parameters vary depending on the background used:
$\Delta \Gamma=\pm0.04$ and $\Delta F_{\rm 3-10\,keV}=\pm 3\times 10^{-14}\rm \ erg\ cm^{-2}\ s^{-1}$
(standard deviations; 1$\sigma$).
We combine these systematic uncertainties in quadrature and report them in Table~\ref{ta:ta1} (second errors).

\begin{figure*}
\centering
\begin{tabular}{cc}
\includegraphics[width=3.3 in]{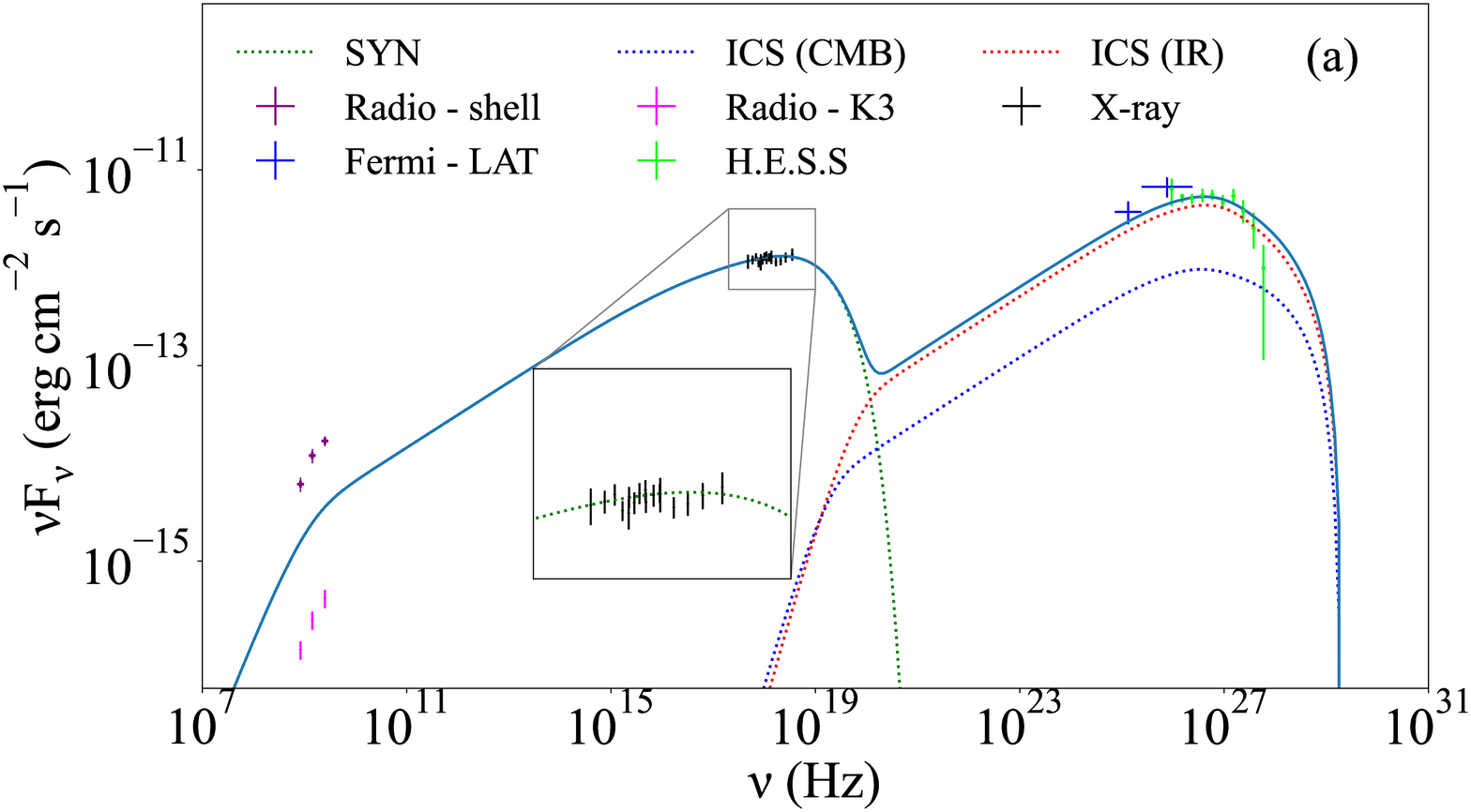} &
\includegraphics[width=3.3 in]{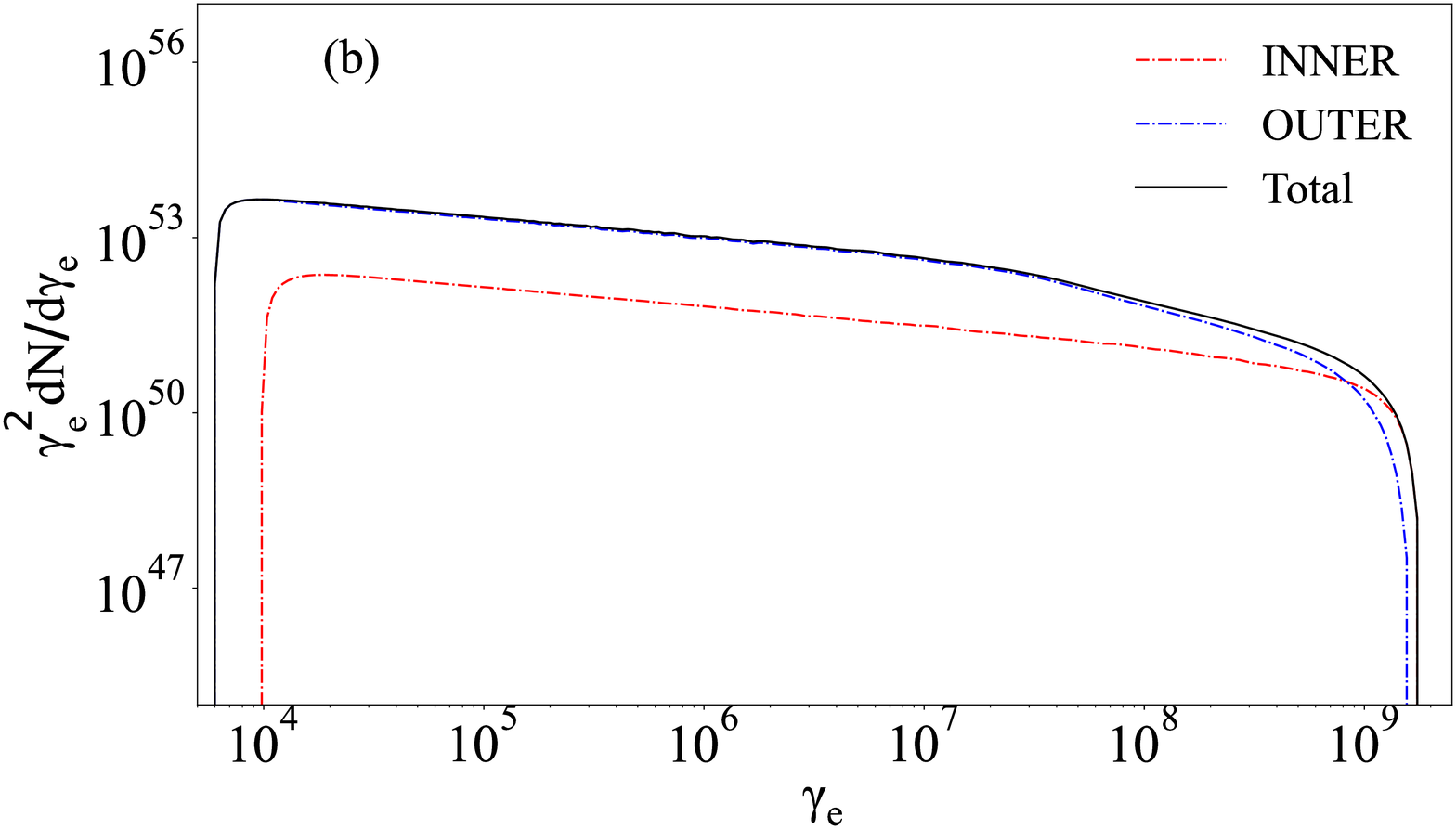} \\
\vspace{-3 mm}
\includegraphics[width=3.3 in]{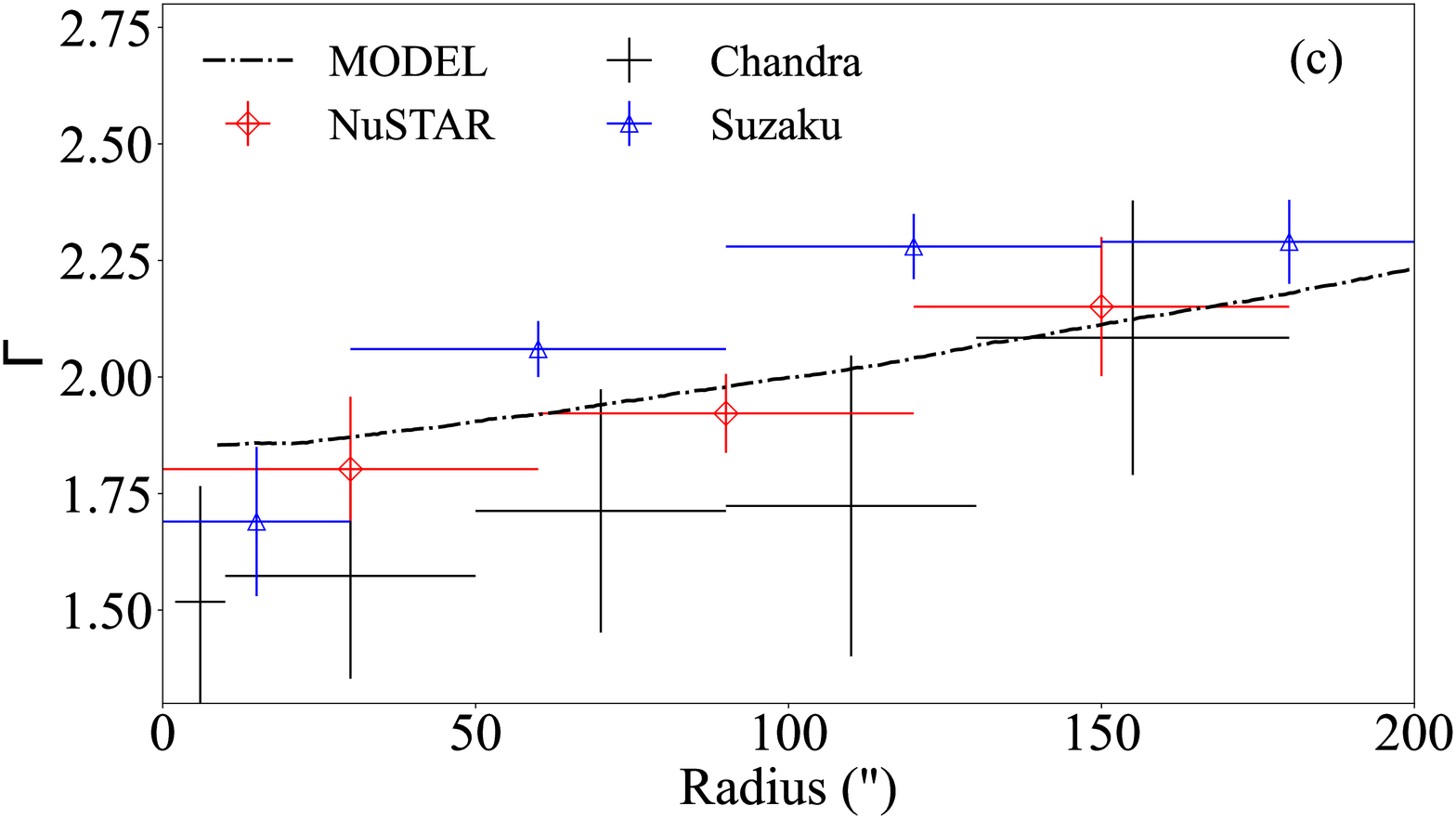} &
\includegraphics[width=3.3 in]{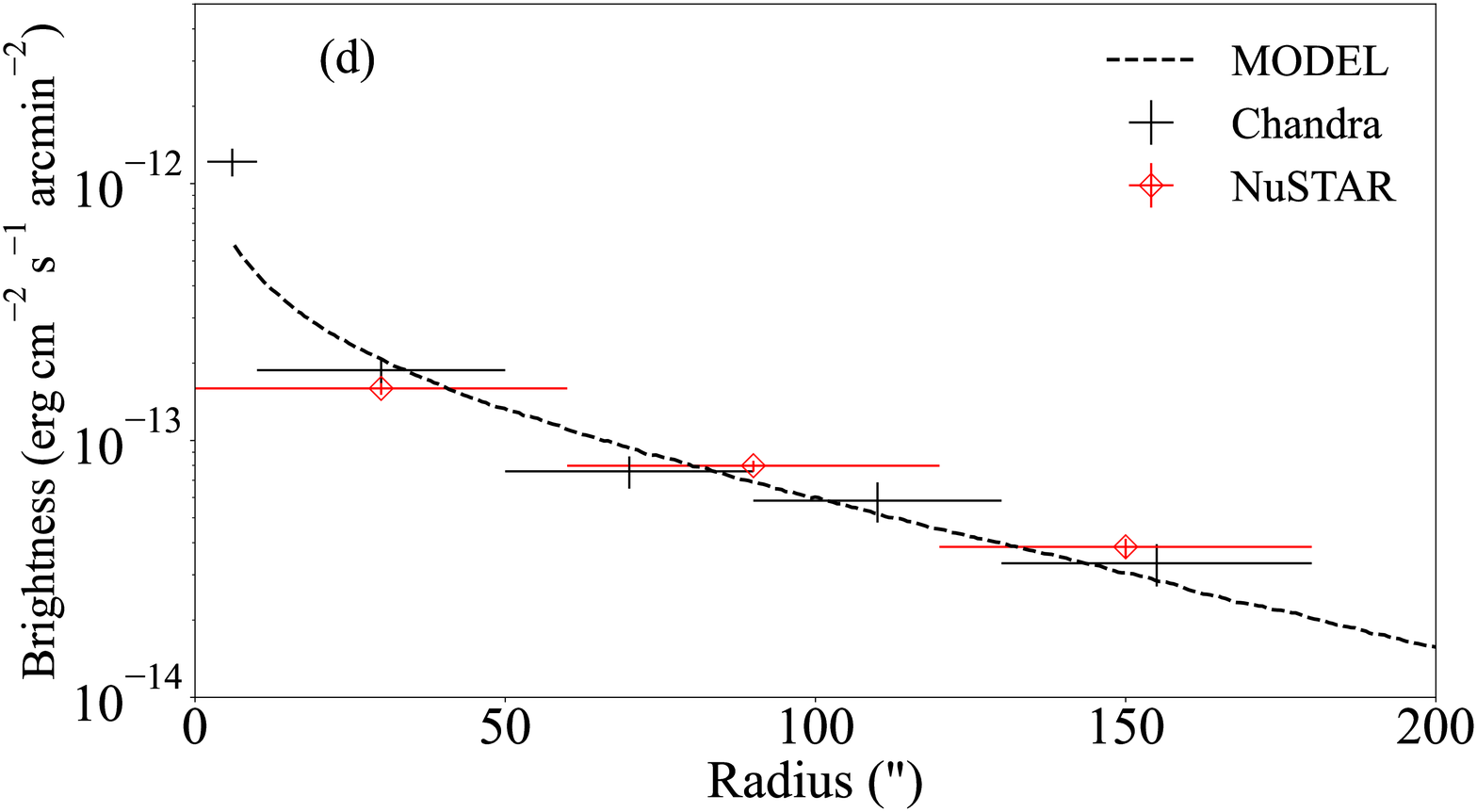} \\
\end{tabular}
\hspace{-5mm}
\figcaption{Broadband SED, radial profiles of X-ray photon index and brightness measured
for the K3 PWN, and an optimized multi-zone emission model.
({\it a}) Spatially integrated broadband SED (Section~\ref{sec:sec4_1}) data and the
optimized model. The radio points are the K3 excess \citep[pink;][]{Roberts1999} and
the shell emission \citep[purple;][]{VanEtten2010}; we take these as the lower and upper limit,
respectively, as was done by \citet{VanEtten2010}. The X-ray points are our measurements of
the PWN emission within the $R=2.5'$ region, and the blue and green points show
$>$20\,GeV Fermi-LAT data \citep[][]{fermi4fgldr3} and 
the H.E.S.S. measurements taken from \citet{hessrabbit2006}, respectively.
The curves are model-computed emission components: green for the synchrotron, red and blue for the ICS from
IR and CMB seeds, and blue solid for the summed model.
({\it b}) Model-computed particle distributions in the inner ($r<100''$; red) and
the outer ($100''<r<150''$; blue) regions, and their sum (black),
({\it c}--{\it d}) radial profiles of X-ray photon index (c) and brightness (d).
In panel (c), we also show Suzaku measurements \citep[][]{Kishishita2012} in blue for reference.
\label{fig:fig5}
}
\vspace{0mm}
\end{figure*}

To measure the PWN spectrum accurately, we jointly fit the 0.5--10\,keV Chandra and
3--20\,keV NuSTAR spectra of the PWN (Fig.~\ref{fig:fig4} and Fig.~\ref{fig:fig5} a).
Note that we modeled the Chandra spectrum with a single power law and the NuSTAR spectra with two power laws with the second power law representing the
pulsar emission (fixed to the Chandra pulsar model; Table~\ref{ta:ta1}).
The best-fit parameters for the PWN were inferred to be $N_{\rm H}=(4.6\pm0.3)\times 10^{22}\rm \ cm^{-2}$,
$\Gamma=1.98\pm0.07$ and $F_{\rm 3-10\,keV}=(1.50\pm0.07)\times10^{-13}\rm \ erg\ cm^{-2}\ s^{-1}$.
The cross-normalization factors for the NuSTAR FPMA and FPMB spectra
with respect to the Chandra spectrum were slightly higher
than, but consistent with 1 at the $\lapp 2\sigma$ level.
We also achieved an acceptable fit to the PWN spectra with a broken power law ($\chi^2$/dof=324/315).
A comparison of the broken power-law and power-law fits found an $f$-test probability
of 0.8, suggesting no significant evidence for spectral curvature or cut-off.

Note again that we used the {\tt angr} abundances for the quoted $N_{\rm H}$ value, to 
compare to previously
published studies and be consistent throughout. Using the newer {\tt Wilms} abundances \citep[][]{wam00} gives, as expected, a higher column density of $(6.94\pm0.55)\times 10^{22}\rm \ cm^{-2}$ but does not change the other spectral parameters.

Because the results may still be influenced by Chandra's background, we checked them
with the 10 Chandra background selections and found 1-$\sigma$ variations of the
parameters to be $\Delta N_{\rm H}=\pm0.4\times 10^{22}\rm \ cm^{-2}$, $\Delta \Gamma=\pm0.03$
and $\Delta F_{\rm 3-10\,keV}=\pm 2\times10^{-14}\rm \ erg\ cm^{-2}\ s^{-1}$.
The joint-fit results also vary depending on the NuSTAR background simulation and the pulsar model.
Uncertainties due to the former and the latter were estimated to be $\Delta N_{\rm H}=\pm 10^{21}\rm \ cm^{-2}$,
$\Delta \Gamma=\pm 0.04$, and $\Delta F_{\rm 3-10\,keV}=\pm 10^{-14}\rm \ erg\ cm^{-2}\ s^{-1}$
(1$\sigma$), and $\Delta N_{\rm H}=\pm 10^{21}\rm \ cm^{-2}$,
$\Delta \Gamma=^{+0.06}_{-0.03}$, and $\Delta F_{\rm 3-10\,keV}=\pm 10^{-14}\rm \ erg\ cm^{-2}\ s^{-1}$
(min/max), respectively.
These systematic uncertainties are combined in quadrature and reported in Table~\ref{ta:ta1}.

\subsubsection{Chandra spectra of the sub-structures in the PWN}
\label{sec:sec2_4_3} 
We measured the X-ray spectra of the sub-structures found in the Chandra image: the torus, jet,
and northern tail (Fig.~\ref{fig:fig2}). The spectra of the other structures (e.g., knots) were difficult to
measure due to the paucity of counts.
To extract spectra, we used elliptical regions with sizes of $10''\times 7''$ (excluding a $R=2''$ region around the pulsar; Fig.~\ref{fig:fig2}a), $7''\times 17''$ (Fig.~\ref{fig:fig2}b), and $2'\times 6'$ (Fig.~\ref{fig:fig2}c) for the torus, jet, and tail, respectively. Background spectra were extracted in the vicinity of the source regions. We then grouped the spectra to have at least 20 counts per bin and fit the spectra with absorbed power-law models holding $N_{\rm H}$ fixed at $4.6\times 10^{22}\rm \ cm^{-2}$. The results are presented in Table~\ref{ta:ta1}.

\subsubsection{Spatially resolved spectra of the PWN}
\label{sec:sec2_4_4}
The spectrum of the K3 PWN has been reported to soften with increasing distance from
the pulsar from $\Gamma\approx1.7$--1.8 to 2.1--2.3 \citep[][]{VanEtten2010,Kishishita2012}.
Our image analysis result (Section~\ref{sec:sec2_3_2}) supported this softening.
We further investigate the spectral softening with the Chandra and NuSTAR data by measuring
spatially resolved spectra of the PWN.
Because the source is not very bright, we use large extraction regions.

In the Chandra data, the source spectra were extracted from five annular regions:
$R=2$--$10''$, 10--50$''$, 50--90$''$, 90--130$''$, and 130--180$''$ centered on J1420.
Background regions were chosen using the procedure described above (based on the blank-sky data and area fraction of
the source region in the chips; Section~\ref{sec:sec2_4_2}).
We grouped the spectra to have a minimum of 30 counts per bin and 
jointly fit them with an absorbed power-law model having an unlinked  $\Gamma$ for each spectrum and a common (and frozen) $N_{\rm H}$ of $4.6\times 10^{22}\rm \ cm^{-2}$.
We fit the data with 10 different background regions to estimate the systematic uncertainties (1$\sigma$ variations of the best-fit parameters).
The min-max variation of $\Gamma$ was small in the innermost region ($\Delta \Gamma\le 0.1$)
but as large as $\Delta \Gamma \approx 0.6$ in outer regions.
The flux variation was modest ($<$10\%).
Because the $\Gamma$ variations are substantial compared to the statistical uncertainties,
we averaged the best-fit parameters and added their 1$\sigma$ variations to the statistical
uncertainties in quadrature; these are displayed in Figure~\ref{fig:fig5} c and d.
The Chandra constraint on the $\Gamma$ profile is poor, and thus we rely on the Suzaku \citep[][]{Kishishita2012} and NuSTAR measurements (see below)
of the $\Gamma$ profile for our SED modeling (Section~\ref{sec:sec4_2}).

For the NuSTAR data, we used a $R=1'$ circle and two annular regions with a width of $1'$ each,
covering a $R=3'$ circular region. Background spectra were generated
with the {\tt nuskybgd} simulations. As noted above, the off-pulse intervals are
narrow, and thus we used all the phases and jointly modeled the pulsar emission.
Its influence on the PWN spectrum decreases rapidly with distance from the pulsar
due to the PSF effect, which was taken into account by reducing the normalization factor
of the pulsar model according to the enclosed energy fraction of the
PSF \citep[e.g.,][]{amwb+14} in each zone.
We grouped the spectra to have at least 30 counts per spectral bin and
jointly fit the spectra with power-law models having separate $\Gamma$s and a common $N_{\rm H}$ (frozen at $4.6\times 10^{22}\rm \ cm^{-2}$).
The fit was acceptable with $\chi^2$/dof=385/371, and the resulting best-fit parameters
are presented in Figure~\ref{fig:fig5} c and d. Note that systematic uncertainties
(e.g., varying the pulsar model and {\tt nuskybgd} simulations; Section~\ref{sec:sec2_4_2})
were added to the statistical uncertainties in quadrature.
For the NuSTAR data, a model with a common $\Gamma$ for the three spectra also yielded a good fit with $\chi^2$/dof=391/373. An $f$-test comparison of the common and separate $\Gamma$ models found $p$=0.07, which 
is weakly suggestive of a spatial variation of $\Gamma$, possibly supporting the previous measurements of the spectral softening in K3 \citep[][]{VanEtten2010,Kishishita2012}.
The NuSTAR-measured $\Gamma$ values are slightly smaller than those measured by Suzaku,
possibly 
because of the different PSF profiles and/or a cross-calibration issue \citep[][]{mhma+15}.

\section{Modeling of the PWN emission}
\label{sec:sec4}
\subsection{Broadband SED data}
\label{sec:sec4_1}
To investigate the properties of particle flow in the K3 PWN using a multi-zone
emission model, we collected radio and TeV SED measurements from the
literature \citep[][]{Roberts1999,VanEtten2010,hessrabbit2006} and took Fermi-LAT data from the 4FGL DR-3 catalog \citep[][]{fermi4fgldr3}. We then added them to our X-ray measurements to construct a broadband SED of the source (Fig.~\ref{fig:fig5}).
Note that \citet{VanEtten2010} regarded the radio
measurements as upper limits for the radio emission of the PWN because an association between 
the X-ray PWN and the apparent radio shell were not established.
We also take the K3 excess and the shell emission as the lower and upper limits, respectively,
for our SED analysis.

\begin{figure}
\centering
\includegraphics[width=3 in]{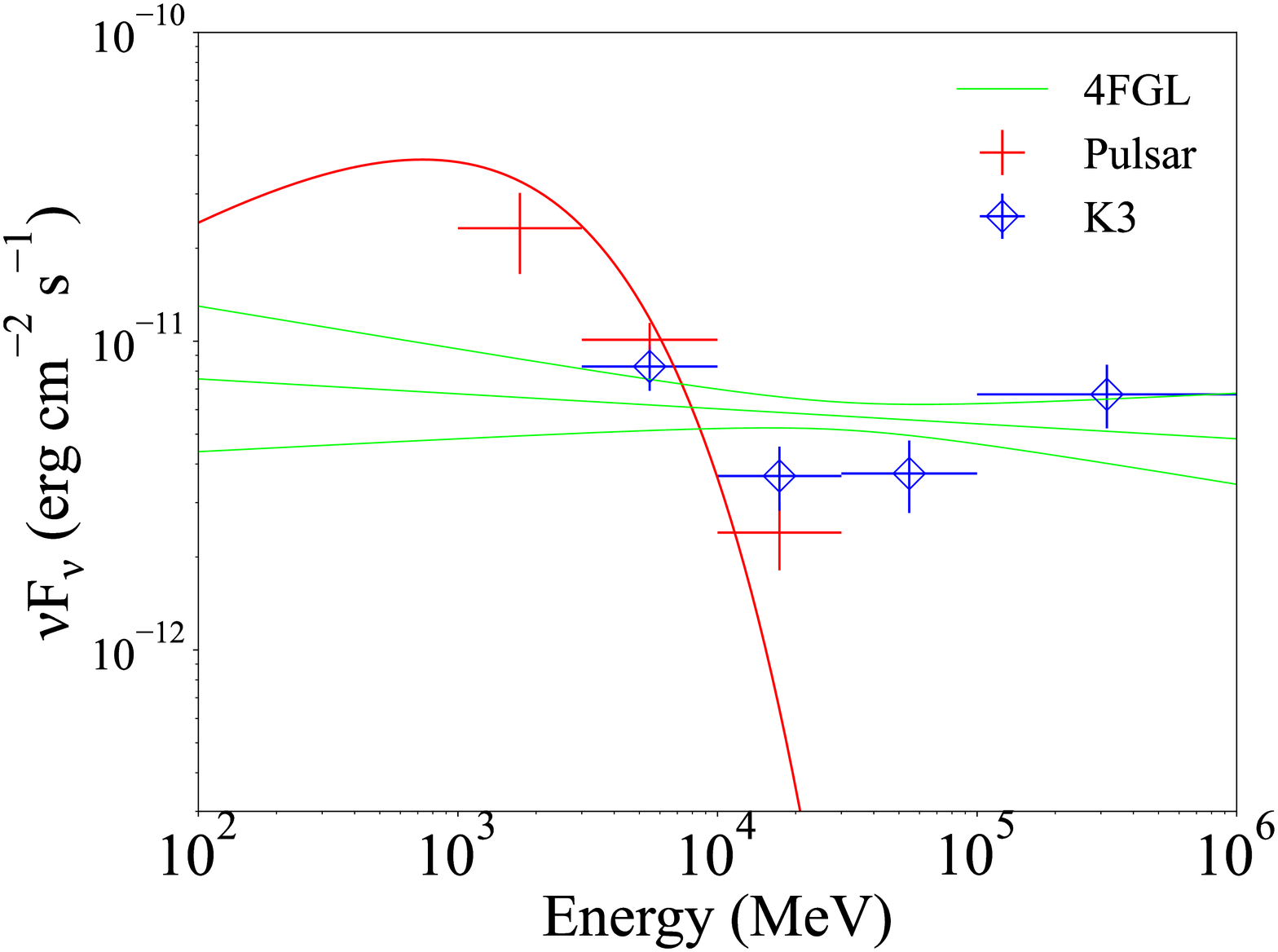} 
\figcaption{Fermi-LAT SEDs in the 100\,MeV--1\,TeV band (taken from the 4FGL DR-3 catalog).
The pulsar's SED points and the {\tt PLSuperExpCutoff4} \citep[][]{fermi4fgldr3} model are shown in red, and
the blue points are the measured SED of the PWN 4FGL~J1420.3$-$6046e.
The catalog model for the PWN is displayed in a green band.
\label{fig:fig6}}
\vspace{0mm}
\end{figure}

The K3 PWN was detected as an extended source (4FGL~J1420.3$-$6046e) in the Fermi-LAT
4FGL DR-2 and DR-3 catalogs \citep[e.g., {\tt gll\_psc\_v28.fit};][]{fermi4fgl,fermi4fgldr3},
and the bright gamma-ray pulsar J1420 is within the PWN. The nebular emission was modeled by 
a power law with photon index $\Gamma=2.00\pm0.13$
and $\Gamma=2.05\pm0.09$ (green in Fig.~\ref{fig:fig6}) in the DR-2 and DR-3 catalog, respectively.
In our inspection of the LAT SEDs reported in the catalogs, we found that low-energy ($\lapp$20\,GeV) flux points of the PWN exhibit a rapidly falling trend with increasing energy
(two blue points at $\lapp$20\,GeV in Fig.~\ref{fig:fig6}).
This trend is reminiscent of the exponentially cut-off
pulsar spectrum (red in Fig.~\ref{fig:fig6}). 
We suspect that some of the high-energy SED tail of
the gamma-ray pulsar 4FGL~J1420.0$-$6048 (LAT counterpart of J1420)
was ascribed to the PWN model in the catalogs. This
speculation is supported by the fact that the PWN is
flagged as a `variable' source probably because it was confused with J1420
as noted by \citet{fermi4fgl}. So we use only the $>$20\,GeV SED points.

The broadband SED in the radio to VHE band is presented in Figure~\ref{fig:fig5} a. Note that we adopted X-ray SED data from within $R=2.5'$ (Section~\ref{sec:sec2_4_2}) and thus integrate our computed X-ray SEDs only out to that distance (see below).

\subsection{Multi-zone PWN emission modeling}
\label{sec:sec4_2}
We applied our phenomenological multi-zone emission model \citep[][]{KimS2020}
to the broadband SED and the measured X-ray profiles of photon index and brightness of K3.
In the model, the pulsar supplies electrons and magnetic field $B$ to the PWN.
The spin-down power $\dot E_{\rm SD}$ of the pulsar goes into energies of the particles ($\dot E_e\equiv \eta_e \dot E_{\rm SD}$) and $B$ ($\dot E_B\equiv \eta_B \dot E_{\rm SD}$) in the PWN, and the pulsar's gamma-ray radiation ($L_\gamma\equiv \eta_\gamma \dot E_{\rm SD}$).
As was done in \citet{Gelfand2009}, we assume $\dot E_{\rm SD}$ evolves over time following 
$\dot E_{\rm SD}(t)=L_0\left (1+\frac{t}{\tau_0}\right )^{-\frac{n+1}{n-1}},$
where $\tau_0=2\tau_c/(n-1)-t_{\rm age}$ and
$n$ is the braking index (assumed to be 3 in this work).
We further assume a constant $\eta_e$, so $\dot E_e$ also evolves with the same time dependence as $\dot E_{\rm SD}$.
Electrons with a
power-law energy distribution, $dN_e/d\gamma_e dt = N_0 \gamma_e^{-p_1}$,
are injected at the termination shock and flow in the PWN via advection and diffusion.

The emission model assumes a spherical flow whose properties, magnetic field strength $B$,
flow speed $V_{\rm flow}$, and diffusion coefficient $D$, are assumed
to be power laws: $B(r)=B_0 (r/R_{\rm TS})^{\alpha_{\rm B}}$, $V_{\rm flow}(r)=V_0(r/R_{\rm TS})^{\alpha_{\rm V}}$,
and $D=D_0(\gamma_e/10^9)(B/100\mu\rm G)^{-1}$, where $R_{\rm TS}$ is the distance from the pulsar to the termination shock. These properties are time-independent in our model, but in reality are likely to change with time because of the time-dependent energy injection $\dot E_{\rm SD}$ (see Section~\ref{sec:sec5_4}).
We estimate $\eta_B$ by comparing the time-integrated $\dot E_{\rm SD}$ to the space-integrated magnetic energy density: $\eta_B \int_{0} ^{t_{\rm age}} \dot E_{\rm SD} dt=4\pi \int_{R_{\rm TS}} ^{R_{\rm PWN}} (B^2/8\pi) r^2 dr$. $\eta_\gamma$ is computed by comparing the measured $L_\gamma$ to the present-day $\dot E_{\rm SD}(t_{\rm age})$. We then require $\eta_e +\eta_B +\eta_\gamma \lapp 1$. 

Particle evolution (in space and energy) is computed in 1000 spatial zones and
$10^4$ $\gamma_e$ bins and for $10^5$ time steps, considering radiative and adiabatic losses \citep[e.g.,][]{tc12}.
We compute the synchrotron and ICS spectra of the particles in each of the emission zones.
For the ICS seed photons, we use the CMB ($T_{\rm CMB}=2.7$\,K with energy density $u_{\rm CMB}=0.26\rm \ eV\ cm^{-3}$)
and IR ($T_{\rm IR}$ with energy density $u_{\rm IR}$) radiation; the parameters for the latter are adjusted to match the VHE SED.
We project the computed emission onto the tangent plane of the observer,
construct spatially integrated and resolved SEDs,
and compare them with measurements (e.g., Fig.~\ref{fig:fig5}).
For the comparison, we integrate the model emission over the projected areas
appropriate for the SED measurements (i.e., 2.5$'$ for X-ray and 0.12$^\circ$ for VHE). Note that the Fermi LAT measured the size of the K3 PWN to be 0.123$^\circ$ using a disk model, whereas H.E.S.S. measured it to be 0.08$^\circ$ using a Gaussian model (1$\sigma$ width). In this work, we adopt the LAT-measured size, but a different size could be accommodated by our model with a change of $u_{\rm IR}$.

Since there are many covariant parameters, it is infeasible to determine all of them with the given measurements
of the broadband SED and radial profiles.
We, therefore, make some assumptions about the PWN flow.
For $B$, we assume transverse configuration \citep[e.g.,][]{Bucciantini2022} and magnetic flux conservation:
i.e., $\alpha_V+\alpha_B=-1$ \citep[][]{r09}.
We further assume an age of $t_{\rm age}=9$\,kyr based on the gamma-ray-to-X-ray
flux ratio \citep[see][]{Kargaltsev2013}, and $R_{\rm TS}=0.14$\,pc
and $R_{\rm PWN}=10.75$\,pc according to the torus size ($\sim$5$''$) and the VHE size of the PWN ($0.12^\circ$),
respectively, for an assumed distance of $d=5.6$\,kpc.
As the X-ray emission seems to be confined within a certain radius from the pulsar,
the PWN properties ($B$ and $V_{\rm flow}$) may change abruptly at the boundary. We verified
that the model-computed emissions did not alter much whether or not we used such abrupt changes
in our modeling since the predicted brightness is very low in the outer regions (e.g., Fig.~\ref{fig:fig5} d).

\begin{table}[t]
\vspace{-0.0in}
\begin{center}
\caption{Parameters for the multizone SED model}
\label{ta:ta2}
\vspace{-0.05in}
\scriptsize{
\begin{tabular}{lcc} \hline\hline
Parameter  & Symbol   & Value   \\ \hline
Spin-down power       & $\dot E_{\rm SD}$              & $10^{37}\rm \ erg\ s^{-1}$     \\
Characteristic age of the pulsar & $\tau_{\rm c}$   & 13000\,yr        \\
Age of the PWN        & $t_{\rm age}$              & 9000\,yr        \\
Size of the PWN       & $R_{\rm pwn}$  & 10.75\,pc        \\
Radius of termination shock & $R_{\rm TS}$  & 0.14\,pc        \\
Distance to the PWN  & $d$  & 5.6\,kpc        \\ \hline
Index for the particle distribution   & $p_1$            & 2.33          \\
Minimum Lorentz factor  & $\gamma_{e,\rm min}$  & $10^{4.45}$        \\
Maximum Lorentz factor  & $\gamma_{e,\rm max}$ & $10^{9.24}$        \\
Magnetic field        & $B_0$      & 5.1$\mu$G        \\
Magnetic index        & $\alpha_B$   & $-$0.06 \\
Flow speed            & $V_0$        & 0.14$c$ \\
Speed index           & $\alpha_V$    & $-$0.94 \\
Diffusion coefficient & $D_0$    & $1.2\times 10^{26}\rm \ cm^2 \ s^{-1}$ \\
Energy fraction injected into particles   &  $\eta_e$      &  0.9        \\
Energy fraction injected into $B$ field        & $\eta_{B}$    & 0.007      \\ \hline
Temperature of IR seeds  & $T_{\rm IR}$  & 10\,K        \\
Energy density of IR seeds  & $u_{\rm IR}$  & 1.7$\rm \ eV \ cm^{-3}$   \\ 
CMB temperature  & $T_{\rm CMB}$  & 2.7\,K   \\
CMB energy density & $u_{\rm CMB}$  & 0.26$\rm \ eV \ cm^{-3}$   \\ \hline
\end{tabular}}
\end{center}
\vspace{-0.5 mm}
\end{table}

Some of the model parameters can be roughly estimated based on the observations.
As Klein-Nishina suppression for electron scattering from IR and CMB seed photons is expected 
at $\gapp$100\,TeV,
the ICS emission of K3 occurs mostly in the Thomson regime.
Then, for the observed ICS to synchrotron SED peak ratio of $\sim$4 and an assumed $u_{\rm IR}$ in the Galaxy of $\approx1\rm \ eV\ cm^{-3}$,
$\frac{B^2}{8\pi}\approx \frac{u_{\rm IR} + u_{\rm CMB}}{4}$ gives $B\approx 4\rm\, \mu$G.
For this $B$, $\approx$20\,keV photons observed from K3 imply a maximum $\gamma_e$ of $\approx 5\times 10^8$.
The uncooled spectrum of the electrons can be directly inferred from the IR-to-optical (synchrotron)
and the LAT (ICS) SEDs. With the lack of IR/optical measurements, a LAT photon index of $\approx$1.7 measured by differencing the two $>$20\,GeV LAT points (Fig.~\ref{fig:fig6}) implies a $p_1=2\Gamma - 1$ of $\approx$2.4.

Using the aforementioned estimates as a guide, we optimized the model to match the
broadband SED and radial profiles of the X-ray brightness and photon index.
An optimized model is displayed in Figure~\ref{fig:fig5}
with parameters given in Table~\ref{ta:ta2}.
We found that particle transport is dominated by advection which
carries the particles to $\sim$12\,pc over the age of 9\,kyr as compared to
the diffusion length $2\sqrt{Dt_{\rm age}}$ of $\approx$2\,pc and $\approx$5\,pc for
VHE (e.g., $\gamma_e\approx 10^7$) and X-ray emitting (e.g., $\gamma_e\approx 10^8$) electrons, respectively.
The diffusion length scale for the highest-energy electrons ($\gamma_e\approx 10^9$)
is comparable to the advection length scale.

In our model, the most energetic particles in K3 have energies of $\approx 10^{15}$\,eV and cool substantially
via synchrotron emission over the pulsar's age (Fig.~\ref{fig:fig5} b). 
Therefore, the synchrotron spectra in outer regions are softer than those in inner regions (i.e., spectral softening by synchrotron burn-off effects).
The computed radial profiles of $\Gamma$ and brightness
depend sensitively on the flow properties (e.g., $B$, $\alpha_B$, $D_0$, etc.)
and thus provide crucial information on the model parameters.
For example, a model with a large $B$ or $\alpha_B$ has difficulty accommodating the large X-ray/VHE extension of the source ($\approx$10\,pc) because particles would have 
cooled efficiently via synchrotron radiation in the inner zones, giving fainter emission  
at large distances.
In contrast, a model with a smaller $B$ would make the $\Gamma$ and brightness profiles flatter. Moreover, such a low-$B$ model requires more particles (i.e., larger $\eta_e$) to match the observed SED. Then the observed $\dot E_{\rm SD}=10^{37}\rm \ erg\ s^{-1}$ of J1420 may be insufficient to provide the required amount of particles; i.e., $\eta_e+\eta_B+\eta_\gamma>1$ for a measured $\eta_\gamma$ of $\approx$7\% for J1420.

The VHE emission primarily arises from the ICS from the IR seeds (Fig.~\ref{fig:fig5} a)
by electrons with energies of $\approx$10\,TeV.
The H.E.S.S. flux ($>$\,TeV) is lower than (but within uncertainties of) the LAT flux ($<$1\,TeV),
possibly indicating a cross-calibration issue. We adjusted our model to match the
H.E.S.S. measurements because they have smaller uncertainties.
Our model slightly overpredicts the $>$10\,TeV SED points, especially the highest-energy one. It may be due to statistical fluctuations, but if real, the highest-energy measurement is hard to explain with our simple model. As the ICS emission at the highest energies
is mainly produced in inner regions (see Section~\ref{sec:sec5_4}), reducing the IR seeds in those regions may help to reconcile the small discrepancy.
Note that the $u_{\rm IR}$ value we used is higher than the Galactic average dust emission \citep[e.g.,][]{Vernetto2016} but is similar to values inferred for other PWNe \citep[e.g.,][]{Torres2014} as PWNe are in crowded regions.

\section{Discussion}
\label{sec:sec5}

\subsection{The pulsar J1420 and the K3 PWN}
\label{sec:sec5_1}
It was firmly established by a Chandra image \citep[][]{Ng2005}
and detection of radio and gamma-ray pulsations \citep[][]{DAmico2001,Weltevrede2010}
that J1420 is associated with the K3 PWN.
The energetic pulsar ($\dot E_{\rm SD}=10^{37}\rm \ erg\ s^{-1}$) has
a power-law spectrum with a photon index $\Gamma=0.7$ 
and a flux $F_{3-10\rm keV}=1.3\times 10^{-13}\rm \ erg\ cm^{-2}\ s^{-1}$ which corresponds
to a 3--10\,keV X-ray luminosity of $\approx5\times 10^{32}\rm \ erg\ s^{-1}$
for the assumed distance of 5.6\,kpc.
For its $f=14.66$\,Hz and $\dot f=-1.77\times 10^{-11}\rm \ Hz\ s^{-2}$,
the measured emission properties of J1420 appear to accord with correlations
seen in rotation-powered pulsars (RPPs)
between temporal and emission properties: e.g., correlations of $\Gamma$ and X-ray luminosity $L_{\rm X,PSR}$
with $\dot E_{\rm SD}$, $\tau_c$, $P=1/f$, and $\dot P=-\dot f/f^2$ \citep[e.g.,][]{Li2008}.
The measured properties of J1420 and K3 are also in accordance with correlations found between
properties of other RPP/PWN systems; e.g., $\Gamma$ and
X-ray luminosity $L_{\rm X,pwn}$ of PWNe are
correlated with $P$, $\dot P$, $\tau_c$, $\dot E_{\rm SD}$, $\Gamma$ and $L_{\rm X, PSR}$ of the pulsars \citep[][]{Gotthelf2003,Li2008}. So we conclude that J1420/K3 is not significantly different from other RPP/PWN systems.

Using the XMM-Newton and NuSTAR data, we unambiguously detected the X-ray pulsations
of J1420 up to 30\,keV (Section~\ref{sec:sec2_2}),
which supports the idea that J1420 is a soft $\gamma$-ray pulsar as suggested by \citet{kh15}.
The X-ray pulse profile of J1420, with a sharp peak and a broad bump, is similar to that of
another soft $\gamma$-ray pulsar PSR~J1418$-$6058.
Intriguingly, these two pulsars seem to exhibit
similar gamma-ray pulse profiles having a smaller peak followed by a bridge and a brighter peak.\footnote{https://www.slac.stanford.edu/$\sim$kerrm/fermi\_pulsar\_timing/}
In the case of PSR~J1418$-$6058, the sharp X-ray peak phase-aligns with the smaller gamma-ray peak \citep[][]{Kim2020}, but
the X-ray and gamma-ray phase alignment is unclear in the case of J1420
because the existing LAT timing solution does not cover the X-ray epochs.
As the relative phasing can provide clues to the spin orientation of the pulsar, further
LAT timing studies are warranted.

\subsection{The PWN morphology}
\label{sec:sec5_2}
In the Chandra data, we identified several small-scale X-ray features: knots and a torus-jet structure.
Although `Knot 1' may be a point source overlapping with
the torus by chance, no detection of an IR or optical counterpart in the 2MASS and USNO
catalogs within $<5''$ and a rather broad spatial distribution observed for Knot1 (compared to other point sources with similar counts)
suggest that it may be a diffuse PWN.
If so, we may speculate based on its location (near the jet base)
that Knot 1 may correspond to a dynamical feature near the jet base
similar to those seen in the Crab nebula \citep[e.g.,][]{Hester2002}.
Knot 2 is also close to the jet but not well aligned with it. Hence, the nature of Knot 2 remains uncertain.

\begin{figure}
\centering
\includegraphics[width=3.2 in]{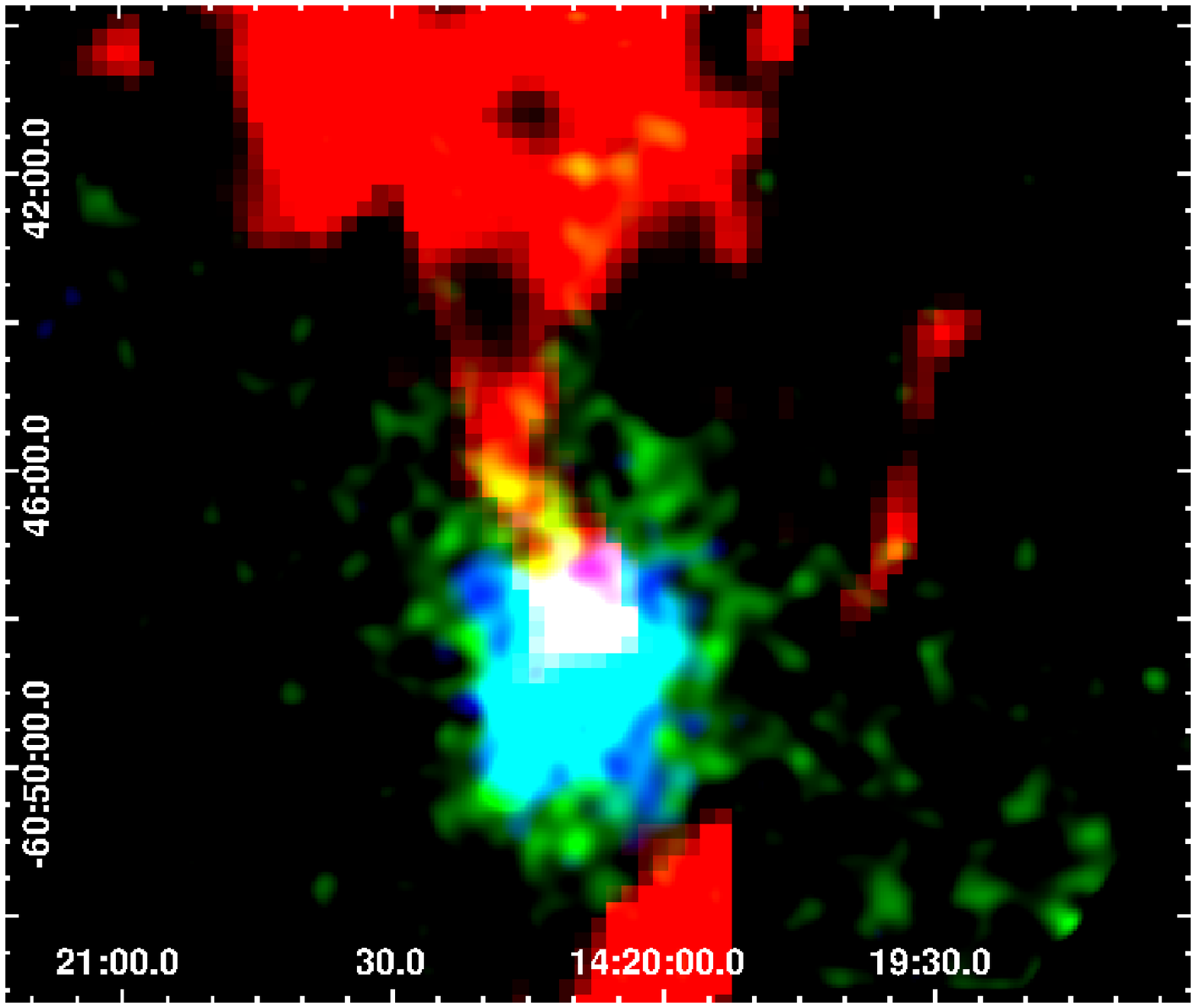}\\
\includegraphics[width=3. in]{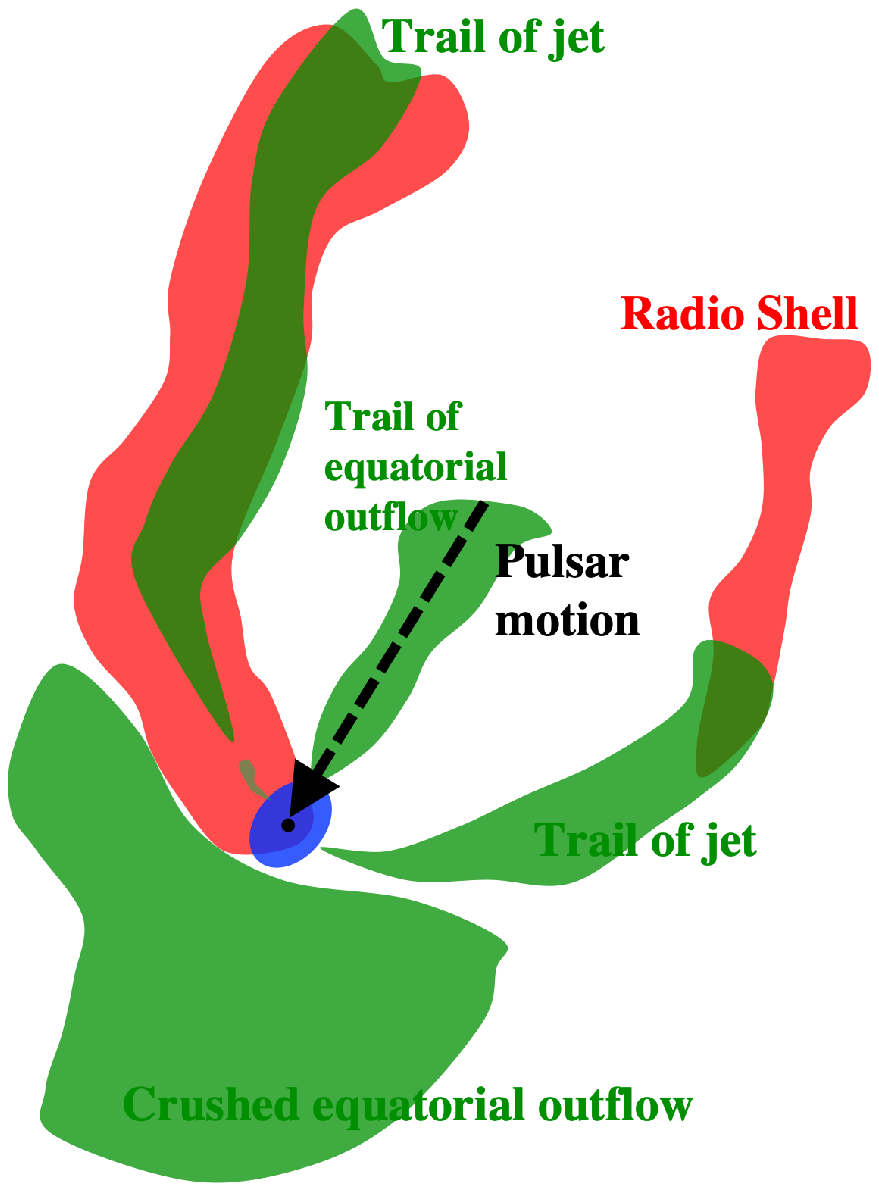}
\figcaption{A false-color image of the source (top) and a schematic (bottom) that shows our speculation on the emission components.
({\it Top}) A radio and X-ray image of the PWN: red for the SUMSS 843\,GHz,
green for the Chandra 1--7\,keV, and blue for the NuSTAR 3--20\,keV image (see Figs.~\ref{fig:fig2} and \ref{fig:fig3}).
({\it Bottom}) Structures (tails and southern bright region) seen in the Chandra image are
presented in green, and the apparent radio shell structure seen
in the SUMSS image is depicted in red. The black dotted arrow shows a suggested pulsar's trajectory \citep[e.g.,][]{VanEtten2010}.
\label{fig:fig7}
}
\vspace{0mm}
\end{figure}

The torus-jet structure we found (Fig.~\ref{fig:fig2} a) can be interpreted
as the termination shock \citep[e.g.,][]{Ng2005}
and a collimated jet. The torus radius of $\sim5''$ corresponds to 0.14\,pc for the
assumed distance of 5.6\,kpc and is consistent with the size of a termination shock
formed by pressure balance \citep[e.g.,][]{Kargaltsev2008}. So the
termination shock interpretation of the torus is plausible.
Then the spin orientation of J1420 may be inferred by a torus model \citep[][]{nr04};
the directions of the torus and the jet suggest that the position angle of the spin axis
is $\approx40^\circ$ east from the north (i.e., along the jet), and an aspect ratio of $\sim$0.6--0.7 of the torus morphology
may suggest that the spin-axis is $\sim$35--45$^\circ$ into or out of the tangent plane if the torus is a ring as
is observed in the Crab Nebula. Further, the fact that a southern jet is not detected may imply that the northern jet is pointing out of the plane.
These crude estimates need to be updated
by more precise torus modeling \citep[e.g.,][]{nr04},
which can be further checked with
pulse profile modeling \citep[e.g.,][]{hma98,rw10}.

The long northern tail detected by Suzaku was interpreted as a trail of the pulsar
which had been born 3$'$ northwest at the center of an apparent radio shell \citep[][see also Fig.~\ref{fig:fig7}]{VanEtten2010}.
In this scenario, the three tails in the Chandra image (Fig.~\ref{fig:fig2} c) may be related to the pulsar's polar and equatorial outflows (Fig.~\ref{fig:fig7} bottom) analogous to those observed in Geminga \citep[][]{Posselt2017}. In this picture, the much more prominent (top) tail (`Tail' in Fig.~\ref{fig:fig2} c) is the bent, originally approaching polar jet, the middle tail is the equatorial outflow, and the bottom tail is the bent counter jet (Fig.~\ref{fig:fig7} bottom).
The putative radio shell discovered by \citet{VanEtten2010} might be
compressing the PWN in the south-east, and the narrow inner jets (Fig.~\ref{fig:fig2} middle)
might have been blocked by the shell, turning into the broad tails over the course of the pulsar's motion (e.g., Fig.~\ref{fig:fig7} bottom).
The compressed southern region would have stronger magnetic field and hence brighter synchrotron emission.
Then particles in the south would cool rapidly via the synchrotron process,
and their VHE emission would be extended in the other direction,
perhaps accounting for the offset VHE emission.
Alternatively, if the pulsar's motion is very small, one can speculate that the
irregular morphology might be caused by inhomogeneities in the ambient medium
\citep[e.g., asymmetric reverse shock interaction;][]{VanEtten2010}.
These scenarios can be distinguished by a measurement of the pulsar's proper motion.
In the bent outflow scenario, the pulsar is expected to move $0.014''
(13{\rm kyr}/t_{\rm age})\rm \ yr^{-1}$ in the south-east direction;
a future high-resolution observatory \citep[e.g., Lynx or AXIS;][]{Gaskin2019,Mushotzky2019} might
be able to detect this proper motion with a $\gapp$20-yr baseline (e.g., in 2030s).

It is intriguing to note that no significant radio emission coinciding with the bright X-ray emission is seen in the southern region. In contrast to K3, other young or middle-aged PWNe exhibit morphological similarity in the radio and X-ray bands \citep[][]{Ng2005,Matheson2010}. It is unclear what causes the lack of radio emission in K3, but it implies that both $B$ and particle energies are high in the south of K3 so that the electrons emit synchrotron radiation above the radio band. These may be related to a reverse shock interaction. If so, we may speculate that the reverse shock interaction started rather recently, and thus the electrons have not cooled sufficiently to produce significant radio emission;
i.e., the synchrotron peak frequency $\nu_{\rm SY}\propto B\gamma_{e,\rm min}^2$ lies above the $\sim$GHz radio band.

\subsection{SED modeling}
\label{sec:sec5_4}
Although the multi-zone model reproduced the observed data broadly, the parameter values are not
well constrained 
due to model assumptions and parameter covariance.
In particular, it is clear that the K3 PWN has some sub-structures (Fig.~\ref{fig:fig2}) and is not spherically symmetric (Fig.~\ref{fig:fig7}).
Moreover, the particle flow in the PWN may be very complex due to pulsar motion and reverse shock
compression \citep[e.g.,][Section~\ref{sec:sec5_2}]{VanEtten2010}. Our model, with the assumptions of spherical flow,
homogeneous IR seed photons and smooth particle distribution,
does not take into account the complex morphology and flows in the bent outflow and/or compressed regions,
which may introduce some inaccuracy to the parameter estimations;
e.g., the $\alpha_B + \alpha_V=-1$ relation may not be valid.
In addition, the PWN properties $B$ and $V_{\rm flow}$ were assumed to be constant in time; these might be higher at early times as $\dot E_{\rm SD}$ was larger. This effect has some influence on the old particles in outer regions and would make their spectra softer than was predicted by our model. However, the broadband SED might be little affected since the X-ray emission from the outer zones is weak (e.g., Fig.~\ref{fig:fig5}).
Note also that the complex parameter covariance of the SED model was not thoroughly investigated,
and there are likely other sets of parameters that may explain the data equally well.
Hence, the parameters reported in Table~\ref{ta:ta2} may represent only the average properties
and need to be taken with caution.
Nonetheless, we discuss below a few intriguing parameters inferred from the modeling.

The spectral softening demands a reasonably strong $B$, but
if it is too strong the model-predicted brightness profile drops too rapidly with increasing distance
to match the measured X-ray profiles.
Our model-inferred $B_0\approx 5\rm \,\mu$G is comparable to those
inferred from a one-zone \citep[$3$--$4\mu\rm G$;][]{Kishishita2012,Zhu2018}
or a two-zone model \citep[$8\mu\rm G$;][]{VanEtten2010}.
We further found that the magnetic index $\alpha_B$ is also small ($-0.06$), meaning that
$B$ in the PWN does not vary much spatially.
The particle motion is mainly driven by the bulk motion $V_{\rm flow}$ and for the assumed
age of 9\,kyr, $V_0=0.14c$ and $\alpha_V=-0.94$ ($V \sim 1/r$) are sufficient to carry the particles to
the outer boundary of the PWN.
These model parameters as well as the hard X-ray emission of K3 suggest that electrons are accelerated to very high energies ($\approx$1\,PeV) in the system.

The inferred parameters for the K3 PWN are generally similar to those for the
middle-aged Rabbit PWN associated with PSR~J1418$-$6058, which has a similar $\tau_c$ (10\,kyr) to J1420's, but the diffusion coefficients $D_0$
differ by an order of magnitude because their observed radial profiles of $\Gamma$
are very different (Park et~al. submitted). X-ray spectral softening was insignificant in 
the Rabbit PWN and thus a large value of $D_0$ was required by the model to homogenize the effects of
synchrotron cooling. An alternative explanation for the lack of softening, negligible diffusion, required an
unreasonably low value of $B$ to avoid a sharp spectral cutoff before the edge of the nebula (see below).
For the $D_0$ values inferred for the two PWNe, the primary particle transport mechanisms 
were also inferred to differ: advection-dominant for K3 and diffusion-dominant for Rabbit. Note, however, that for the highest-energy particles ($\gamma_e\approx 10^9$) the effect of diffusion is comparable to that of advection even in K3.
While the parameters inferred by our phenomenological model may not
be accurate, the distinctions (i.e., $\Gamma$ profiles) observed between
the two PWNe convincingly suggest that their diffusion properties differ.
This may be related to the evolution of the PWNe, and their interaction with the
putative supernova remnants and/or the ambient medium. These need to be
investigated with more physically-motivated evolution models.

\begin{figure}
\centering
\includegraphics[width=3.3 in]{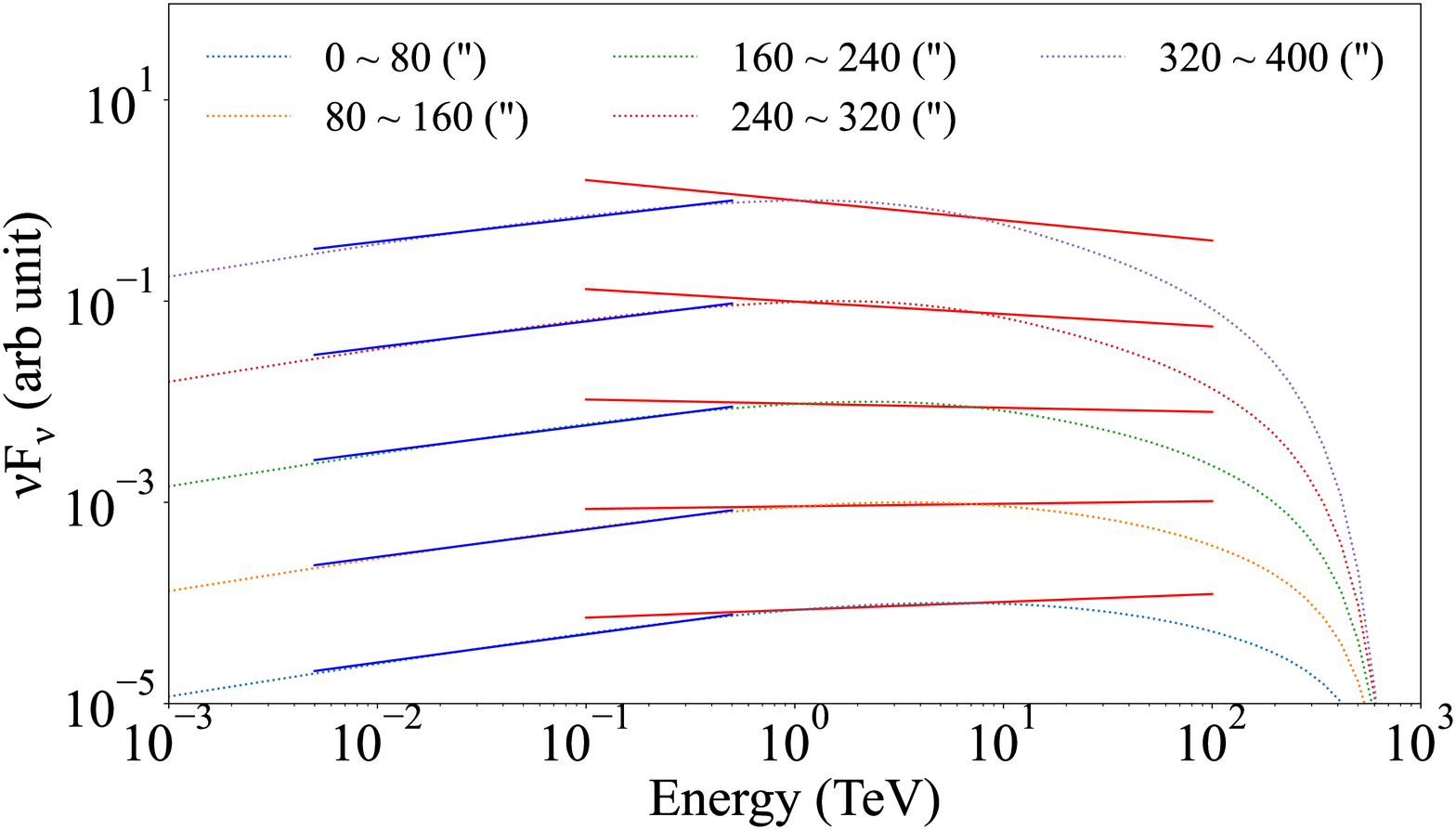} \\
\includegraphics[width=3.3 in]{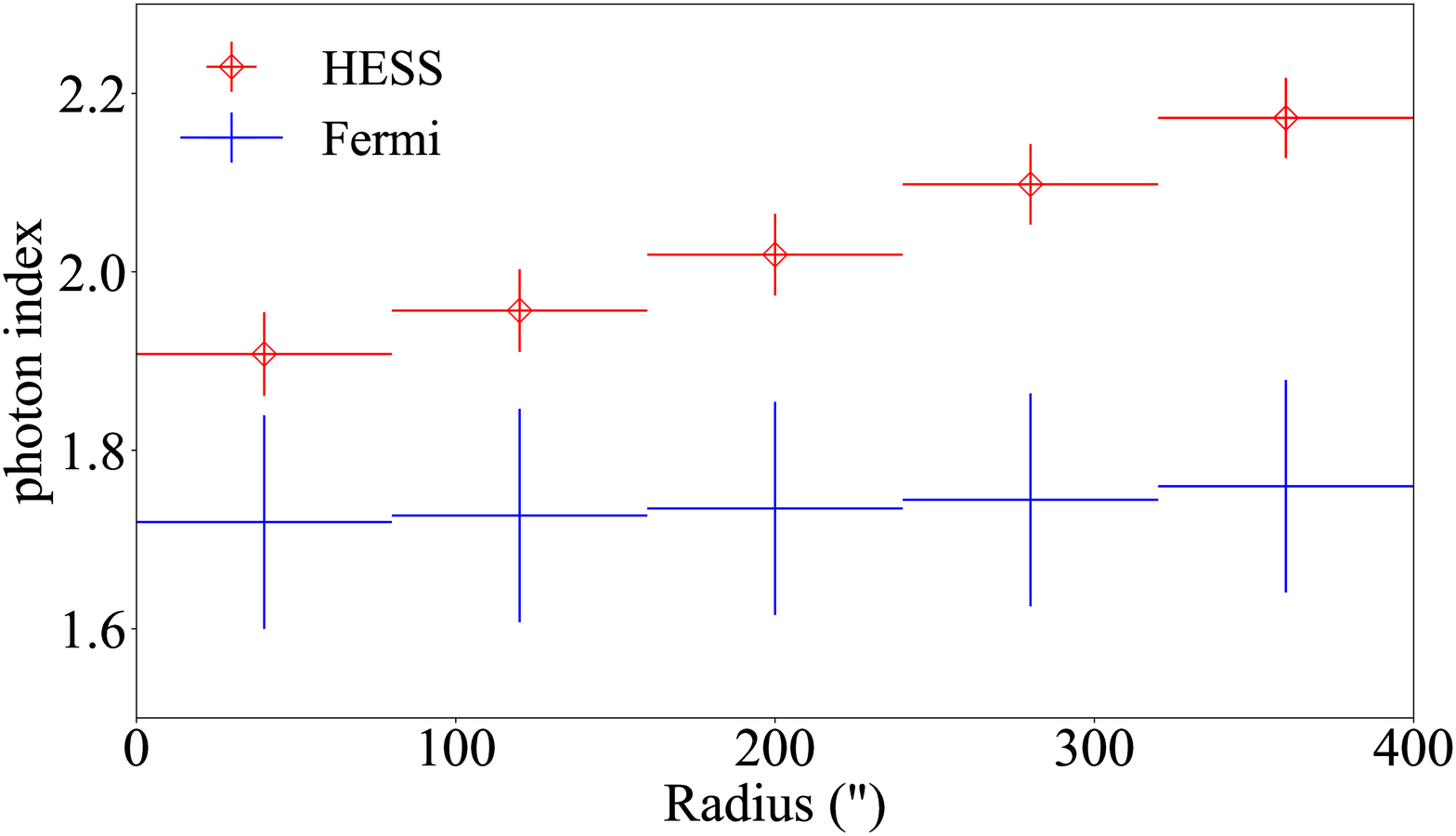}
\figcaption{Spatially resolved VHE SED models and effective photon indices computed with
the model fluxes. ({\it Top}) VHE SEDs from an inner (bottom) to an outer (top) region.
The SEDs are displaced in the y-axis for legibility. The straight lines
drawn on the SEDs are effective power-law fits to
simulated SED data that are constructed using the model-computed fluxes and relative
uncertainties on the spatially integrated LAT and H.E.S.S. SED measurements.
({\it Bottom}) Photon indices of the effective power-law models in the LAT (blue) and the
H.E.S.S. (red) bands. See text for more detail.
\label{fig:fig8}
}
\vspace{0mm}
\end{figure}

Although the particle transport in K3 is dominated by advection,
the mild and steady increase of $\Gamma$ with distance is different
from those expected in purely advective flows.
In these cases, all the particles at a certain radius have the same age
(cooling time). So in inner regions, where the cooling break
is above X-ray energies, the synchrotron spectral index in the X-ray band does not change radially.
As the particles propagate outward in the later evolutionary stage, the cooling break
moves to lower energies, eventually to X-rays. From that radius the X-ray spectrum
softens rapidly with increasing radius (age). This trend is almost universally observed
in 1D pure-advection models \citep[e.g.,][]{Reynolds2003}.
Such a trend can turn into a steadily increasing one (observed in K3) if particles
of different ages can mix. This can be achieved by non-radial backflows
that bring older (outer) particles back into the inner regions and/or by
diffusion. Although the former may play a role in K3 (e.g., in the southern region), the details of the backflows are not well known and thus difficult to incorporate into our spherically symmetric model.
Hence, we relied on particle diffusion in this work.
For the X-ray emitting electrons, the diffusion length scale in K3 was estimated
to be $\sim$5\,pc (or longer for the highest-energy electrons)
which is sufficient to mix particles of various ages (c.f., 12\,pc for advection).
Note that \citet{tc12} demonstrated that $\Gamma$ profiles flatten at large distances in models with reflecting outer boundary conditions probably by producing backflows,
whereas $\Gamma$ steadily increases with distance in the absence of the backflow (i.e., transmitting outer boundary).
The Suzaku measurement of the $\Gamma$ profile of K3 seems to exhibit a flattening (Fig.~\ref{fig:fig5}),
but a better measurement with finer radial bins is needed to confirm it.

Our model predicts a cooling break at $\gamma_e\approx 10^7$. Because particles with that Lorentz factor
emit ICS radiation at $\approx$1\,TeV, spatial variation of emission 
below (i.e., LAT band) and above it (i.e., H.E.S.S. band) are expected to be substantially different
as is demonstrated in Figure~\ref{fig:fig8} which displays model-predicted VHE SEDs and effective photon indices in the LAT and H.E.S.S. bands.
Note that this is only for demonstration, and the actual uncertainties on the spatially resolved
SED points would vary depending on the instrument and exposure.
The slope of the LAT SED does not vary spatially because the $<$1\,TeV emission is
produced by uncooled electrons, whereas the cooling effect is clearly visible in the TeV
band. Our model predicts a significant spectral softening in the H.E.S.S. band
at a similar level to that seen in the X-ray band (Fig.~\ref{fig:fig5} c).
This can be tested with deeper VHE observations of the source
\citep[e.g., by CTA;][]{CTA2011}.

\section{Summary}
\label{sec:sec6}
We analyzed broadband X-ray data to characterize the emission properties of
J1420 and K3. Below we summarized our findings.
\begin{itemize}
\item We confidently detected the X-ray pulsations of J1420, using NuSTAR and XMM-Newton,
and measured its pulse profile accurately.
\item We identified sub-structures of the K3 PWN in the Chandra image:
two knots, a torus-jet structure, and large-scale tails.
\item Using NuSTAR images, we found a hint of a spectral softening with increasing distance
from J1420. This is in accord with the previous measurements of the spectral softening in K3
\citep[][]{VanEtten2010,Kishishita2012}.
\item Our multi-zone emission modeling suggested that particles are accelerated
to very high energy ($\approx$1\,PeV), the nebular magnetic field is low ($B \sim 5\mu$G), and
the particles are transported primarily by advection in the K3 PWN.
\end{itemize}

The high-quality X-ray data, especially those taken with NuSTAR  thanks to its hard X-ray coverage,
were useful
for our study of the particle properties in the K3 PWN at the highest energies. A spectral cut-off
was not detected to $\sim$20\,keV, implying that there may exist even higher-energy
particles in the PWN; these can be probed by future observatories operating above the NuSTAR band \citep[e.g., FORCE, HEX-P and COSI;][]{Nakazawa2018,Madsen2019,Tomsick2019}. Comparisons with other middle-aged PWNe
can help to further our understanding of PWN physics \citep[e.g.,][]{Kargaltsev2013,Mori2021}. For example,
X-ray spectral softening has been observed in K3, but not
in the Rabbit PWN; this was attributed to different levels of diffusion
in our model. However, there may be other reasons (different evolution or environments)
that the current data and models do not capture. Further observational and theoretical
efforts are needed.
Importantly, a measurement of the pulsar proper motion is needed to confirm the evolutionary scenario of this complex system.

\bigskip
\bigskip

\acknowledgments
This work used data from the NuSTAR mission, a project led by the California Institute of Technology,
managed by the Jet Propulsion Laboratory, and funded by NASA. We made use of the NuSTAR Data
Analysis Software (NuSTARDAS) jointly developed by the ASI Science Data Center (ASDC, Italy)
and the California Institute of Technology (USA).
This research was supported by Basic Science Research Program through
the National Research Foundation of Korea (NRF)
funded by the Ministry of Science, ICT \& Future Planning (NRF-2022R1F1A1063468).
Support for this work was partially provided by NASA through NuSTAR Cycle 6
Guest Observer Program grant NNH19ZDA001N.
SSH acknowledges support from the Natural Sciences and Engineering Research Council
of Canada (NSERC) through the Discovery Grants and Canada Research Chairs programs
and from the Canadian Space Agency (CSA).
We thank the referee for detailed comments that helped improve the clarity of the paper.
\vspace{5mm}
\facilities{CXO, XMM-Newton, NuSTAR}
\software{HEAsoft \citep[v6.29;][]{heasarc2014}, CIAO \citep[v4.14;][]{fmab+06},
XMM-SAS \citep[20211130\_0941;][]{xmmsas17}, XSPEC \citep[v12.12;][]{a96}}


\bibliographystyle{apj}
\bibliography{ms.bbl}

\end{document}